\documentclass[twocolumn,english,showpacs,preprintnumbers,amsmath,amssymb,floatfix]{revtex4-1}

\usepackage[T1]{fontenc}
\usepackage[latin9]{inputenc}
\usepackage{color}
\usepackage{array}
\usepackage{amstext}
\usepackage{graphicx}
\usepackage{esint}
\usepackage{rotating}
\usepackage[font=small,labelfont=bf]{caption}
\usepackage{hyperref}

\makeatletter

\providecommand{\tabularnewline}{\\}

\@ifundefined{textcolor}{}
{%
 \definecolor{BLACK}{gray}{0}
 \definecolor{WHITE}{gray}{1}
 \definecolor{RED}{rgb}{1,0,0}
 \definecolor{GREEN}{rgb}{0,1,0}
 \definecolor{BLUE}{rgb}{0,0,1}
 \definecolor{CYAN}{cmyk}{1,0,0,0}
 \definecolor{MAGENTA}{cmyk}{0,1,0,0}
 \definecolor{YELLOW}{cmyk}{0,0,1,0}
 }

\@ifundefined{definecolor}
 {\usepackage{color}}{}
\@ifundefined{definecolor}
 {\usepackage{color}}{}
\makeatother
\usepackage{babel}

\begin{document}


\title { Next-to-next-to-leading order QCD analysis of spin-dependent parton distribution functions and their uncertainties: Jacobi polynomials approach }

\author { F. Taghavi-Shahri$^{1}$ }
\email{Corresponding author: Taghavishahri@um.ac.ir}

\author { Hamzeh Khanpour$^{2,3}$ }
\email{Hamzeh.Khanpour@mail.ipm.ir}

\author { S. Atashbar Tehrani$^{4}$ }
\email{Atashbar@ipm.ir}

\author {Z. Alizadeh Yazdi$^{1}$ }
\email{Zahra.Alizadehyazdi@stu.um.ac.ir}

\affiliation {
$^{(1)}$Department of Physics, Ferdowsi University of Mashhad, P.O.Box 1436, Mashhad, Iran   \\
$^{(2)}$Department of Physics, University of Science and Technology of Mazandaran, P.O.Box 48518-78195, Behshahr, Iran  \\
$^{(3)}$School of Particles and Accelerators, Institute for Research in Fundamental Sciences (IPM), P.O.Box 19395-5531, Tehran, Iran\\
$^{(4)}$Independent researcher, P.O. Box 1149-8834413, Tehran, Iran     }

\date{\today}

%
%
\begin{abstract}\label{abstract}
%
We present a first QCD analysis of next-to-next-leading-order (NNLO) contributions of the spin-dependent parton distribution functions (PPDFs) in the nucleon and their uncertainties using the Jacobi polynomial approach. Having the NNLO contributions of the quark-quark and gluon-quark splitting functions in perturbative QCD (Nucl. Phys. B 889 (2014) 351-400), one can obtain the evolution of longitudinally polarized parton densities of hadrons up to NNLO accuracy of QCD. A very large sets of recent and up-to-date experimental data of spin structure functions of the proton $g_1^p$, neutron $g_1^n$, and deuteron $g_1^d$ have been used in this analysis.
The predictions for the NNLO calculations of the polarized parton distribution functions as well as the proton, neutron and deuteron polarized structure functions are compared with the corresponding results of the NLO approximation. We form a mutually consistent set of polarized PDFs due to the inclusion of the most available experimental data including the recently high-precision measurements from {\tt COMPASS16} experiments (Phys. Lett. B 753 (2016) 18-28).
We have performed a careful estimation of the uncertainties using the most common and practical method, the Hessian method, for the polarized PDFs originating from
the experimental errors.
The  proton, neutron and deuteron structure functions and also their first moments, $\Gamma^{\rm p, n, d}$, are in good agreement with the experimental data at small and large momentum fraction of $x$. We will discuss how our knowledge of spin-dependence structure functions can improve at small and large value of $x$ by the recent {\tt COMPASS16} measurements at CERN, the {\tt PHENIX} and {\tt STAR} measurements at RHIC, and at the future proposed colliders such as Electron-Ion collider (EIC).
\end{abstract}


\maketitle

\tableofcontents{}

%
%
\section{Introduction}\label{Introduction}

The "spin crisis" has been a longstanding mystery in high energy particle physics.
In 1987, series of experiments proved that the spins of the quarks are only partially responsible for the proton's overall spin.
Thus for a decade, long search for the missing pieces, or contributors, to a proton's spin have been done.
The key question is how the spin of the nucleon is distributed among its constituent partons. That is, the
determination and understanding the longitudinal spin structure functions of the nucleon $g_1^{\tt N} (x, Q^2) (\rm N = p, n, d)$ and the behavior of spin-dependent parton distribution functions (PPDFs) appeared as an important issue. Recent years have seen increased theoretical interest and setting up experiments towards the better understanding and precise determinations of the polarized nucleon structure function $g_1$, specially in {\tt HERMES}, {\tt COMPASS}, {\tt PHENIX} and {\tt STAR} at a variety of energies.

There are several next-to-leading order (NLO) QCD analysis of the polarized DIS data along with the estimation of their uncertainties in the literature~\cite{Nocera:2014gqa,Nocera:2014uea,Jimenez-Delgado:2014xza,Jimenez-Delgado:2013boa,Nocera:2014gqa,Arbabifar:2013tma,Monfared:2014nta,Borah:2014esa,Khorramian:2010qa,Blumlein:2010rn,Leader:2014uua,Aschenauer:2015ata}. These parton sets differ in the choice of experimental data sets, treatment of heavy quarks, details of the QCD analysis such as higher-twist corrections, the form of the polarized PDFs at input scale and the error propagation.  In this work we provide, for the first time, a next-to-next-to-leading (NNLO) order QCD analysis of polarized parton distribution functions and their uncertainties using all available and up-to-date deeply inelastic scattering data.

The determination of the longitudinal spin structure of the nucleon has attracted considerable theoretical and experimental interests since the surprising {\tt EMC} experimental results showed that the quark contributions to the nucleon spin is very small~\cite{Ashman:1987hv,Ashman:1989ig}. The present knowledge on the longitudinal proton spin structure function, $g^p_1$ originates from measurements of the asymmetry A$^p_1$ in polarized lepton nucleon scattering.
In the recent years, the available DIS data which may be used for the determination of polarized PDFs, has been extended impressively. The most up-to-date longitudinal polarized deep inelastic scattering (DIS) experimental data from the {\tt COMPASS} collaboration~\cite{Adolph:2015saz,Alekseev:2007vi,Alekseev:2009ab,Adolph:2012vj,Adolph:2012ca}, {\tt HERMES} collaboration~\cite{Airapetian:2004zf,Airapetian:2006vy,Airapetian:2010ac}, {\tt PHENIX} collaboration~\cite{Adare:2008qb,Adare:2008aa,Adare:2014hsq,Adare:2010cc,Adare:2010xa} and {\tt STAR} collaboration~\cite{Adamczyk:2013yvv,Adamczyk:2012qj,Adamczyk:2014ozi,Aggarwal:2010vc,Adamczyk:2014xyw} provide very precise information to study the spin structure and quark PDFs inside the nucleon. These data include the semi-inclusive particle production, high-$p_T$ jet production, semi inclusive DIS in fixed target experiments and W$^\pm$ boson production in polarized proton-proton collisions.

The purpose of the following paper is to present for the first time a very good quality of the polarized PDFs at NNLO using the  analysis of available polarized DIS data, taking into account the most recent data from {\tt COMPASS16} measurements~\cite{Adolph:2015saz}. An appealing feature of this QCD analysis of polarized PDFs is that we apply the theoretical predictions at NNLO accuracy in perturbative QCD.
A careful estimation of the uncertainties have been performed using the most common Hessian method for the polarized parton distributions of quarks and gluon originating from the experimental errors. It is shown that the present analysis considerably leads to smaller value of uncertainties in comparison with other polarized PDFs in the literature. The Jacobi polynomials approach is used to facilitate the analysis. A detailed comparison with other available polarized PDFs including {\tt KATAO}~\cite{Khorramian:2010qa}, {\tt BB}~\cite{Bluemlein:2002be},
{\tt GRSV}~\cite{Gluck:2000dy}, {\tt LSS/LSS06}~\cite{Leader:2005ci,Leader:2006xc}, {\tt DNS}~\cite{deFlorian:2005mw}, {\tt AAC04/AAC09}~\cite{Goto:1999by,Hirai:2008aj}, {\tt DSSV08/DSSV10}~\cite{deFlorian:2008mr,deFlorian:2009vb} and the most recent results from {\tt AKS14}~\cite{Arbabifar:2013tma} and {\tt THK14}~\cite{Monfared:2014nta} have been presented. 
Due to recent high precision measurements we also revisit our next-to-leading order QCD analysis of longitudinal spin structure of the nucleon and present an updated, more accurate, version of our polarized PDFs at next-to-leading order of QCD. In order to discuss the fit results, we will concentrate on our NNLO fit. Since the outcome for NLO and NNLO polarized PDFs are slightly different, we will show the results of both QCD fits in some figures.
We also focus on the roles of the NNLO terms on the polarized PDFs determination by comparing the available NLO results with the present NNLO analysis.  Moreover, to establish a meaningful baseline for estimating the impact of higher order corrections and to examine the effect of the change in the NNLO polarized PDFs, we compare our NLO and NNLO analyses which have been extracted from the same DIS data set using exactly the same functional forms for polarized distributions and the same assumptions. 
The main features of our NNLO parametrization of polarized PDFs are worth emphasizing already at this point. The details of the analysis will be present in the next Sections.

The structure of the present paper is as follows:
In Sec.~\ref{PPDFs-analysis-method}, we will turn to the method of the polarized structure function analysis based on the Jacobi polynomials approach.
In Sec.~\ref{global-PPDFs}, we review the input to the global analysis including the data selection and the input parameterizations of the polarized PDFs and deeply inelastic structure functions.
The results of the present polarized PDFs analysis are given in Sec.~\ref{Results-of-PPDFs}. We will study how much a NNLO determination of spin-dependent structure functions would improve our knowledge of polarized parton distribution functions.
In Sec.~\ref{Comparison-with-the-data}, a detailed comparison between the present results and available experimental data are presented. We also have attempted a detailed comparison of our NNLO results with recent results from the literature in this Section.
In Sec.~\ref{RHIC-LHC-era}, we will discuss how our knowledge of spin-dependence structure functions may be improved at small and large value of $x$ by the recent {\tt COMPASS16} measurements at CERN, {\tt PHENIX} and {\tt STAR} measurements at RHIC and at the future proposed colliders such as Electron-Ion collider (EIC).
Finally, we have presented our summary and conclusions in Sec.~\ref{Summary}.

%
%
\section{Polarized PDFs analysis method}\label{PPDFs-analysis-method}
Beyond leading order accuracy of pertuarbative QCD, structure functions are no longer linear combination of quark distributions. At higher order, structure functions are obtained by convoluting the quark and gluon distributions with the corresponding pertuabative coefficient functions. Having the next-to-next-to-leading order (NNLO) contributions of the quark-quark and gluon-quark splitting functions in perturbative QCD~\cite{Moch:2014sna}, one can obtain the evolution of longitudinally polarized parton densities of hadrons up to NNLO order of QCD~\cite{Cafarella:2005zj}.
The NNLO spin-dependent proton structure functions, $g_1^{\rm p} (x, Q^2)$, can be written as a linear combination of polarized parton distribution functions $\Delta q$, $\Delta \bar{q}$ and $\Delta g$ as,
\begin{eqnarray}\label{eq:g1pxspace}
&&g_1^{\rm p} (x, Q^2) = \frac{1}{2} \sum_q e^2_q \left( \Delta q(x, Q^2) + \Delta \bar{q}(x, Q^2)\right)\otimes\nonumber\\ &&\left(1+\frac{\alpha_s(Q^2)}{2 \pi}\Delta C^{(1)}_q+\left(\frac{\alpha_s(Q^2)}{2 \pi}\right)^2\Delta C^{(2)}_q\right)\nonumber\\
&&+\frac{2}{9}\left(\frac{\alpha_s(Q^2)}{2 \pi}\Delta C^{(1)}_g+\left(\frac{\alpha_s(Q^2)}{2 \pi}\right)^2\Delta C^{(2)}_g\right)\otimes\Delta g(x, Q^2)\nonumber\\
\end{eqnarray}
where the $\Delta C_q$ and $\Delta C_g$ are the spin-dependent quark and gluon coefficient functions~\cite{Lampe:1998eu,Zijlstra:1993sh}
The method in which we have employed in the present paper is using the Jacobi polynomials expansion of the polarized structure functions.
The detailed of such analysis based on the Jacobi polynomials are presented in our previous works~\cite{Khorramian:2010qa,Khorramian:2009xz,Khorramian:2008yh,AtashbarTehrani:2007odq} and also other groups\cite{Barker:1982rv,Barker:1983iy,Krivokhizhin:1987rz,Krivokhizhin:1990ct,Chyla:1986eb,Barker:1980wu,Kataev:1997nc,Alekhin:1998df,Kataev:1999bp,Kataev:2001kk,Kataev:2005ci,Leader:1997kw}. In this section we outline a brief review of this method.
The Jacobi polynomials expansion method is one of the simplest and fastest algorithm to reconstruct the structure function from the QCD predictions for its Mellin moments. In this method, one can easily expand the polarized structure functions, $x g_1(x,Q^2)$, in terms of the Jacobi polynomials, $\Theta_{n}^{\alpha, \beta}(x)$, as follows,
\begin{equation}\label{eq:xg1}
x g_{1}(x, Q^2) = x^{\beta} (1 - x)^{\alpha}\ \, \sum_{n = 0}^{N_{max}} a_n(Q^{2}) \, \Theta_n^{\alpha,\beta}(x) \,,
\end{equation}
where $n$ is the order of the expansion terms, $N_{\rm max}$ is the maximum order of the expansion which normally can be set to 7 and 9. The parameters $\alpha$ and $\beta$ are a set of free parameters which normally set to 3 and 0.5, respectively. We have shown in our previous work that by setting the {$N_{\rm max}$ = 7 and 9,  $\alpha$ = 3, $\beta$ = 0.5}, this expansion of the structure function can be achieve to optimal convergence throughout the whole kinematic region constrained by the DIS data. The Q$^2$--dependence of the structure functions are codify in the Jacobi polynomials moments, $a_{n}(Q^{2})$. The $x$-dependence will be provided by the weight function $x^{\beta} (1-x)^{\alpha}$ and the Jacobi polynomials $\Theta_n^{\alpha, \beta}(x)$ which can be written as,
\begin{equation}\label{eq:Jacobi}
\Theta_n^{\alpha, \beta}(x) = \sum_{j = 0}^{n} \, c_j^{(n)}(\alpha, \beta) \, x^j \,,
\end{equation}
where the coefficients $c_j^{(n)}(\alpha, \beta)$ are combinations of Gamma functions in term of $n$, $\alpha$ and $\beta$. The above Jacobi polynomials have to satisfy the following orthogonality relation,
\begin{equation}\label{eq:orthogonality}
\int_0^1 dx \, x^{\beta} (1 - x)^{\alpha} \, \Theta_{k}^{\alpha, \beta}(x) \, \Theta_{l}^{\alpha, \beta}(x) = \delta_{k,l} \,.
\end{equation}
Consequently one can obtain the Jacobi moments, $a_n(Q^2)$, using the above orthogonality relations as,
\begin{eqnarray} \label{eq:aMoment}
a_n(Q^2) & = & \int_0^1 dx \, x g_1(x,Q^2) \, \Theta_k^{\alpha, \beta}(x) \nonumber \\
 & = & \sum_{j=0}^n \, c_j^{(n)}(\alpha, \beta) \, {\cal M} [xg_{1}, j + 2]  \,,
\end{eqnarray}
where the Mellin transform ${\cal {M}} [x g_1, N]$ introduced as,
\begin{eqnarray}\label{eq:Mellin}
{\cal {M}} [x g_1, {\rm N}]  \equiv  \int_0^1 dx \, x^{\rm N-2} \, xg_1 (x, Q^2) \,.
\end{eqnarray}
Finally the polarized structure function $x g_1(x,Q^2)$ can be written as follows,
\begin{eqnarray}\label{eq:g1Jacobi}
x g_1(x, Q^2) & = & x^{\beta}(1 - x)^{\alpha} \, \sum_{n=0}^{N_{max}} \, \Theta_n^{\alpha, \beta}(x) \nonumber \\
&\times & \sum_{j=0}^n \, c_j^{(n)}{(\alpha, \beta)} \, {\cal M}[x g_1, j + 2] \,.
\end{eqnarray}
This method can also be used to construct the proton $F_2^p (x, Q^2)$ and neutron $F_2^n (x, Q^2)$ structure functions~\cite{Khorramian:2009xz,Khorramian:2008yh}.
%
%
\section{Input to the global polarized PDFs fit}\label{global-PPDFs}
%

\subsection{NNLO QCD fits of $g_1$ world data}

We have adopted the following standard parameterizations at the input scale of Q$_0^2$ = 4 GeV$^2$ for the polarized up-valence $x \Delta u_v$,  down-valence $x \Delta d_v$, sea $x \Delta \bar{q}$ and gluon $x \Delta g$ distributions,
\begin{equation}\label{eq:parametrizations}
x \, \Delta q(x, Q_0^2) = {\cal N}_{q} \, \eta_q \, x^{a_q} (1 - x)^{b_q} \, (1 + c_q x) \,,
\end{equation}
where the normalization factors, ${\cal N}_q$, can be determined as,
\begin{equation}\label{eq:Norm}
\frac {1} {{\cal N}_q} = \left (1 + c_q \frac{a_q} {a_q + b_q + 1} \right) \, B \left (a_q, b_q + 1 \right) \,.
\end{equation}
Considering SU(3) flavor symmetry, we have $\Delta \bar{q}  \equiv  \Delta \bar{u} = \Delta \bar{d} = \Delta \bar{s} = \Delta s$. 
Some latest analysis show that including SIDIS data can help to consider light sea-quark decomposition. In the analysis presented in this paper, we wish to study the impact of inclusive DIS data on the determination of NNLO polarized PDFs based on Jacobi polynomials with flavor symmetric light sea distribution. The impact of SIDIS data on the sea quark distributions will be studied in a separate publication in the near future.
The unknown parameters in Eq.~\ref{eq:parametrizations} will be extracted from fit to experimental data. The normalization factors, ${\cal N}_q$, are chosen such that the parameters $\eta_q$ are the first moments $\Delta q_i(x, Q_0^2)$ distributions,
\begin{equation}\label{eq:firstmoments}
\eta_i = \int_0^1 dx \, \Delta q_i  (x, Q_0^2) \,.
\end{equation}
For the $\Delta u_v$ and $\Delta d_v$ polarized valence distributions, the first moment of the corresponding distributions, $\eta_{u_v}$ and $\eta_{d_v}$, will be obtained as,
\begin{eqnarray}\label{eq:constrain}
a_3 & = & \int_0^1 dx \, \Delta q_3 = \eta_{u_v} - \eta_{d_v} = F + D  \,,   \\
a_8 & = & \int_0^1 dx \, \Delta q_8 = \eta_{u_v} + \eta_{d_v} = 3 F - D \,.
\end{eqnarray}
The $a_3$ and $a_8$ are the non-singlet combinations of the first moments of the polarized parton distributions corresponding to $q_3 = ( \Delta u + \Delta \bar{u} ) - (  \Delta d + \Delta \bar{d} )$ and $q_8 = ( \Delta u + \Delta \bar{u} ) - (  \Delta d + \Delta \bar{d} ) - 2 ( \Delta s + \Delta \bar{s}  )$.
The first moments of the polarized valence quark densities introduced in Eq.~\ref{eq:firstmoments} can be related to $F$ and $D$ as measured
in neutron and hyperon $\beta$ decays~\cite{Agashe:2014kda}. These constraints lead to the values of $\eta_{u_v} = 0.928 \pm 0.014$ and $\eta_{d_v} = -0.342 \pm 0.018$ for the $\Delta u_v$ and $\Delta d_v$ polarized valence distributions, respectively.
The Dokshitzer-Gribov-Lipatov-Altarelli-Parisi (DGLAP) evolution equations~\cite{Dokshitzer:1977sg,Gribov:1972ri,Lipatov:1974qm,Altarelli:1977zs} are solved in Mellin space and used in Jacobi polynomial approach. The Mellin transform for the polarized PDFs $q$ are defined as,
\begin{eqnarray}
{\cal M} [\Delta q(x, Q_0^2), {\rm N}] & \equiv & \Delta q({\rm N}, Q_0^2) = \int_0^1 x^{{\rm N} - 1} \, \Delta q(x,Q_0^2) \, dx \nonumber \\
& = & {\cal {N}}_q \eta_q \left(1 + c_q \, \frac{{\rm N} - 1 + a_q}{ {\rm N} + a_q + b_q} \right ) \, \nonumber \\
& \times & B({\rm N} - 1 + a_q, b_q + 1) \,,
\end{eqnarray}
where $q$ is the polarized PDFs as $x \Delta u_v$,  $x \Delta d_v$, $x \Delta \bar{q}$ and $x \Delta g$. In the Mellin space, the twist-2 contributions to the polarized structure functions $g_1 ({\rm N}, Q^2)$ can be written in terms of polarized PDFs, $\Delta q ({\rm N}, Q^2)$,  $\Delta \bar{q} ({\rm N}, Q^2)$ and  $\Delta g ({\rm N}, Q^2)$, and the corresponding coefficient functions $\Delta C^N_i$,
\begin{eqnarray}
&&{\cal M} [g_1^p,  {\rm N}] = \nonumber \\
& = & \frac{1}{2}\sum \limits _q e_q^2 \left\{ \left(1 + \frac{\alpha_s}{2 \pi} \Delta C_q^{(1)}({\rm N}) + \left(\frac{\alpha_s}{2 \pi}\right)^2\Delta C_q^{(2)} (N)\right)\right.\nonumber \\
& \times & [\Delta q(N, Q^2) + \Delta \bar{q} ({\rm N}, Q^2)] \nonumber \\
& + & \left.\frac{2}{9}\left(\frac{\alpha_{s}}{2 \pi} \, \Delta C_g^1 (N) + \left(\frac{\alpha_s}{2 \pi} \right)^2 \, \Delta C_g^2({\rm N})\right) \Delta g({\rm N}, Q^2)\right\} \,. \nonumber\\
\end{eqnarray}

\subsection{Data selection, minimization and error calculation}

The data in which we used in our NNLO polarized PDFs  QCD analysis, are summarized in Table.~\ref{tab:DISdata}.
This table contains the name of the experimental group, the covered kinematic ranges in $x$ and $Q^2$, the number of available DIS data points and the fitted normalization shifts ${\cal{N}}_i$. The data used (465 experimental points) cover the following kinematics region: {$0.0035<x<0.75$ and 1 < Q$^2$ < 96 GeV$^2$}. The global fit reported in the present article incorporates: a wide range of the polarized deeply inelastic scattering lepton-nucleon data on spin structure functions $g_1^p$~\cite{Abe:1998wq,HERM98,Adeva:1998vv,EMCp,E155p,HERMpd,COMP1,Adolph:2015saz}, $g_1^d$~\cite{Abe:1998wq,E155d,Adeva:1998vv,HERMpd,COMP2005,COMP2006} and $g_1^n$~\cite{E142n,HERM98,E154n,HERMn,JLABn2004,JLABn2005}. An important and appealing feature of our NNLO QCD analysis of the polarized
PDFs is that we used the recently published polarized deeply inelastic scattering data from {\tt COMPASS16}~\cite{Adolph:2015saz}. These data sets contain both statistical and systematic errors which added in quadrature. In addition, the normalization errors are generally specified separately.

\begin{table*}[!htbp]
	\caption{ Published data points above Q$^2$ = 1.0 GeV$^2$. Each experiment is given the $x$ and Q$^2$ ranges, the number
		of data points for each given target, and the fitted normalization shifts ${\cal{N}}_i$ (see the text).} \label{tab:DISdata}
	\begin{ruledtabular}
		\begin{tabular}{lccccc}
			\textbf{Experiment} & \textbf{Ref.} & [$x_{\rm min}, x_{\rm max}$]  & \textbf{Q$^2$ range {(}GeV$^2${)}}  & \textbf{Number of data points} &   \textbf{${\cal N}_n$}       \tabularnewline
			\hline\hline
			E143(p)   & \citep{Abe:1998wq}   & [0.031-0.749]   & 1.27-9.52 & 28 & 0.999402403\\
			HERMES(p) & \citep{HERM98}  & [0.028-0.66]    & 1.01-7.36 & 39 & 1.000386936\\
			SMC(p)    & \citep{Adeva:1998vv}    & [0.005-0.480]   & 1.30-58.0 & 12 & 1.000084618\\
			EMC(p)    & \citep{EMCp}     & [0.015-0.466]   & 3.50-29.5 & 10 & 1.010741787\\
			E155      & \citep{E155p}    & [0.015-0.750]   & 1.22-34.72 & 24 & 1.024394035\\
			HERMES06(p) & \citep{HERMpd} & [0.026-0.731]   & 1.12-14.29 & 51 &0.998865500 \\
			COMPASS10(p) & \citep{COMP1} & [0.005-0.568]   & 1.10-62.10 & 15 &0.9942871736\\
			COMPASS16(p) & \cite{Adolph:2015saz} & [0.0035-0.575]   & 1.03-96.1 & 54 &1.0009687352\\
			\multicolumn{1}{c}{$\boldsymbol{g_1^p}$}       &  &  &  &  \textbf{233}  & \\ \hline
			E143(d)  &\citep{Abe:1998wq}    & [0.031-0.749]   & 1.27-9.52    & 28 & 0.9993545553\\
			E155(d)  &\citep{E155d}     & [0.015-0.750]   & 1.22-34.79   & 24 & 1.0001291961\\
			SMC(d)   &\citep{Adeva:1998vv}     & [0.005-0.479]   & 1.30-54.80   & 12 & 0.9999944683\\
			HERMES06(d) & \citep{HERMpd}& []0.026-0.731]   & 1.12-14.29   & 51 & 0.9984082065  \\
			COMPASS05(d)& \citep{COMP2005}& [0.0051-0.4740] & 1.18-47.5   & 12 & 0.9983759396 \\
			COMPASS06(d)& \citep{COMP2006}& [0.0046-0.566] & 1.10-55.3    & 15 & 0.9997379579 \\
			\multicolumn{1}{c}{ $\boldsymbol{g_1^d}$}      &  &  & & \textbf{142} &    \\ \hline
			E142(n)   &\citep{E142n}    & [0.035-0.466]   & 1.10-5.50    & 8 & 0.9989525725 \\
			HERMES(n) &\citep{HERM98}   & [0.033-0.464]   & 1.22-5.25    & 9 & 0.9999732650\\
			E154(n)   &\citep{E154n}    & [0.017-0.564]   & 1.20-15.00   & 17 & 1.0003242284 \\
			HERMES06(n) &\citep{HERMn}  &  [0.026-0.731]  & 1.12-14.29   & 51 & 0.9999512597 \\
			Jlab04(n)&\citep{JLABn2004} & [0.33-0.60]      & 2.71-4.8     & 3 & 0.9997264174 \\
			Jlab05(n)&\citep{JLABn2005} & [0.19-0.20]     &1.13-1.34     & 2 &  1.0002854347 \\
			\multicolumn{1}{c}{$\boldsymbol{g_1^n}$}     &     &  & & \textbf{90} &   \\ \hline\hline \multicolumn{1}{c}{\textbf{Total}}&\multicolumn{5}{c}{~~~~~~~~~~~~~~~~~~~~~~~~~~~~~~~~~~~~~~~~~~~~~~~~~~~~~~\textbf{465}}
			\\
		\end{tabular}
	\end{ruledtabular}
\end{table*}

\begin{figure*}[htb]
	\vspace*{0.5cm}
	\includegraphics[clip,width=0.99\textwidth]{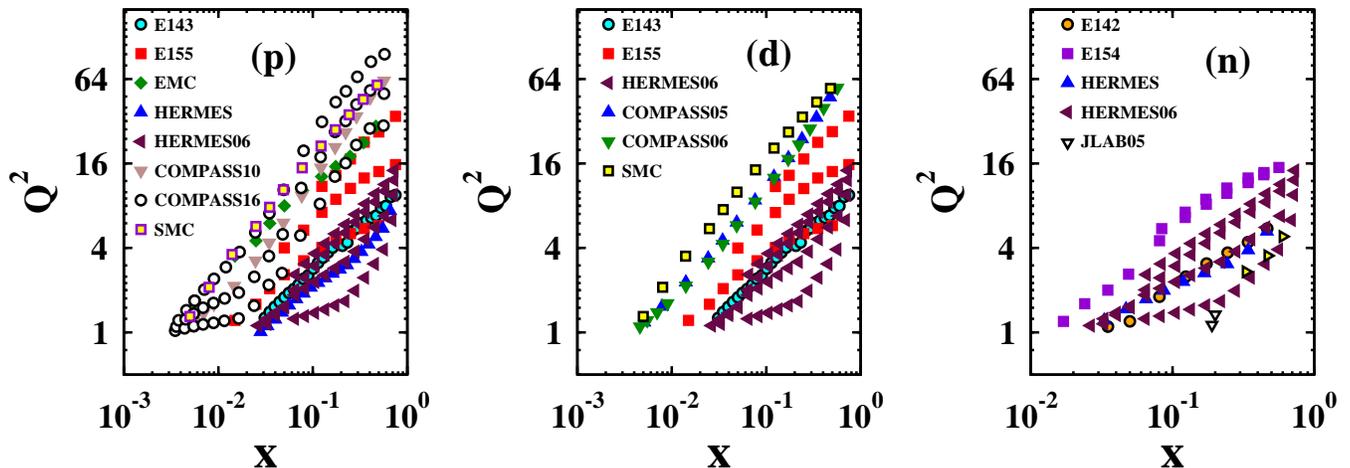}
	\begin{center}
	\caption{ {\small (color online) Nominal coverage of the data sets used in our global fits for proton, neutron and deuteron.} \label{fig:figxQ} }
	\end{center}
\end{figure*}

Nominal coverage of the data sets used in our fits for proton, neutron and deuteron are presented in Fig.~\ref{fig:figxQ}. The plots clearly show that despite remarkable experimental efforts, the kinematical coverage of the present available DIS data being included in analysis of polarized PDFs still rather limited. As we mentioned, the accessed range of momentum fraction $x$ is $0.0035 < x < 0.75$. This coverage can leads to a larger uncertainties for determined polarized PDFs at small $x$. For the gluon distribution, which is the most complicated case for PDF uncertainties and parameterizations, we expected a different treatment at $x < 0.01$ due to the lack of DIS data.

The analysis of $\chi^2$ value and the error calculation based on the Hessian method are applied in the present analysis. For the error calculation, a standard error analysis is needed for the polarized PDFs by taking into account correlations among the parameters. The resulting eigenvector sets of the determined polarized PDFs can be used to propagate uncertainties to any other desired observable. The method to consider the correlations among the uncertainties is discussed in details in Refs.~\cite{Khanpour:2016pph,AtashbarTehrani:2012xh,Martin:2002aw,Pumplin:2001ct,Monfared:2011xf,Arbabifar:2013tma}. Following that, a detailed error analysis has been done using the covariance or Hessian matrix, which can be obtained by running the CERN program library {\tt MINUIT}~\cite{James:1994vla}.
\begin{figure}[htb]
	\vspace*{0.5cm}
	\includegraphics[clip,width=0.45\textwidth]{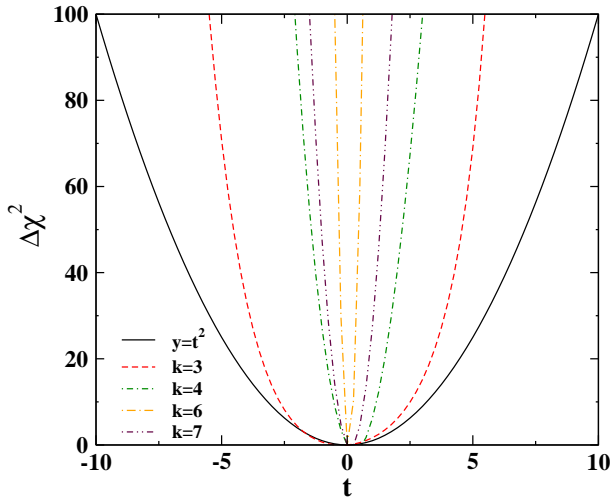}
	\begin{center}
		\caption{ {\small (color online) $\Delta \chi^2$ as a function of $t$ defined in Refs.~\cite{Khanpour:2016pph,AtashbarTehrani:2012xh,Martin:2002aw,Pumplin:2001ct,Monfared:2011xf,Arbabifar:2013tma} for some random sample
				of eigenvectors, $k^{\rm th}$.} \label{fig:chi2} }
	\end{center}
\end{figure}
$\chi_{\rm global}^{2}(\rm p)$ quantifies the goodness of fit to the DIS data for a set of independent parameters $p$ that specifies the polarized PDFs at Q$_0^2$ =  4 GeV$^2$,
\begin{equation}\label{eq:chi2}
\chi_{\rm global}^2 (p) = \sum_{n} w_n \chi_n^2 \,,
\end{equation}

\begin{equation}
\chi_n^2 (p) = \left( \frac{1 -{\cal N}_n }{\Delta{\cal N}_n}\right)^2 + \sum_i \left(\frac{{\cal N}_n  \, g_{1,i}^{\rm exp} - g_{1,i}^{\rm theor} (p) }{{\cal N}_n \, \Delta g_{1,i}^{\rm exp}} \right)^2 \,.
\end{equation}
The minimization of the above $\chi_{\mathrm{\rm global}}^2 (p)$ function is done using the CERN program library {\tt MINUIT}~\cite{James:1994vla}.
In the above equation, $g_{1,i}^{\rm exp}$, $\Delta g_{1,i}^{\rm exp}$, and $g_{1,i}^{\rm theor}$ denote the experimental measurement, the experimental uncertainty
(statistical and systematic combined in quadrature) and the theoretical value for the $i^{\rm th}$ experimental data point, respectively.
${\cal N}_n$ is an overall normalization factor for the data of experiment $n$ and the ${\Delta{\cal N}_n}$ is the experimental normalization uncertainty.
We allow for a relative normalization shift ${\cal N}_n$ between different experimental data sets within uncertainties ${\Delta{\cal N}_n}$ quoted
by the experiments.
We minimize the above $\chi_{\rm global}^2 (p)$ value with the $9$ unknown fit parameters plus an undetermined $\Lambda_{\rm QCD}^{\overline{\rm MS}}$. We find $\chi^{2}/{\rm dof}= 401.92/456=0.881 \ ({\rm NNLO})$ and  $\chi^{2}/{\rm dof}=410.856/456= 0.901 \ ({\rm NLO})$ which yield an acceptable fit to the experimental DIS data. The results show that there is an improvement in the quality of the fit at NNLO.

Some groups such as {\tt NNPDF}~\cite{Nocera:2014gqa,Ball:2013tyh,Ball:2013lla} or {\tt JAM}~\cite{Sato:2016tuz} propose an alternative approach in their analysis for the PDFs uncertainties, based on an iterative Monte Carlo fitting technique that allows a more robust extraction of polarized PDFs with statistically rigorous PDFs uncertainties. What makes {\tt NNPDF} differs from others is using neural networks instead of traditional parametrizations. They have presented a global polarized PDFs determination and achieved a significant improvement in accuracy in the determination of the positive polarized gluon distribution in the medium and small-$x$ region.
In our analysis we utilize, like most of the existing phenomenological spin-dependent PDFs analyses, the standard PDFs fitting technology in which single fits are performed assuming a basic parametric form for the input polarized PDFs. The polarized PDFs errors are then typically computed using the standard error analysis such as Hessian methods proposed by J. Pumplin, D. Stump, Wu-Ki Tung et al. (PST)~\cite{Pumplin:2001ct,Martin:2003sk} which are based on
diagonalization of the matrix of second derivatives for $\chi^2$ (Hessian matrix) near the minimum of $\chi^2$. 
%
%
\begin{table*}
	\renewcommand{\arraystretch}{1.30}
	\centering
	{\footnotesize
		\begin{tabular}{||c||c|c|c|c|c|c|c|c|c||}
			\hline \hline
			& $a_{u_v}$ & $b_{u_v}$ & $a_{d_v}$ & $b_{u_v}$ & $\eta_{\bar{q}}$ & $a_{\bar{q}}$ & $\eta_g$ & $a_g$ & $\alpha_{s}(Q_0^2)$     \\
			
			\hline \hline
			$a_{u_v}$          & 7.7882 $\times 10^{-4}$ &  &  &  &  &  &  & & \\
			\hline
		    $b_{u_v}$          & 1.535 $\times 10^{-3}$  &  4.97 $\times 10^{-2}$&  &  &  &  &  & & \\
			\hline
	         $a_{d_v}$         & 1.279 $\times 10^{-6}$ & 1.904 $\times 10^{-3}$ & 4.841 $\times 10^{-4}$ &  &  &  &  &&  \\
			\hline
		      $b_{d_v}$        &-3.627 $\times 10^{-6}$  & 0.4662 &   9.864 $\times 10^{-5}$&   0.3169              &  &  &  & & \\
			\hline
		$\eta_{\bar{q}}$       & 4.861 $\times 10^{-6}$ &  -1.075 $\times 10^{-2}$  & 1.923 $\times 10^{-6}$  &  1.454 $\times 10^{-3}$ & 2.479 $\times 10^{-3}$     &  &  & & \\
			\hline
	    $a_{\bar{q}}$    	   & 2.000 $\times 10^{-5}$ & -4.713 $\times 10^{-2}$ & 1.642 $\times 10^{-5}$ &  8.317 $\times 10^{-3}$   & -1.302 $\times 10^{-4}$   & 0.6388  &  & & \\
			\hline
	    $\eta_g$               & 2.245 $\times 10^{-3}$ & -5.061  & 3.466 $\times 10^{-3}$ & 0.6800       &  -1.568 $\times 10^{-2}$  & -6.909 $\times 10^{-2}$ & 6.700 $\times 10^{-4}$ & & \\
			\hline
			$a_g$              & 5.969 $\times 10^{-4}$ & -1.345  & -8.430 $\times 10^{-4}$ & 0.1676     & -4.081 $\times 10^{-3}$   &-1.747 $\times 10^{-2}$ & -1.974 $\times 10^{-3}$ &  8.154 $\times 10^{-3}$ & \\
			\hline
	$\alpha_{s}(Q_0^2)$        & -2.137 $\times 10^{-7}$ & -6.011 $\times 10^{-7}$ & 3.375 $\times 10^{-6}$ &1.1561 $\times 10^{-5}$   & -2.557 $\times 10^{-7}$    & 1.878 $\times 10^{-7}$ &  2.660 $\times 10^{-7}$ & -5.021 $\times 10^{-5}$ & 5.76 $\times 10^{-4}$  \\
			\hline \hline
		\end{tabular}
	}
	\caption[]{ The covariance matrix for the $8 + 1$ free parameters in the NNLO fit. }
	\label{covmat-matNNLO}
\end{table*}


\begin{table*}
	\renewcommand{\arraystretch}{1.30}
	\centering
	{\footnotesize
		\begin{tabular}{||c||c|c|c|c|c|c|c|c|c||}
			\hline \hline
			& $a_{u_v}$ & $b_{u_v}$ & $a_{d_v}$ & $b_{u_v}$ & $\eta_{\bar{q}}$ & $a_{\bar{q}}$ & $\eta_g$ & $a_g$ & $\alpha_{s}(Q_0^2)$     \\
			
			\hline \hline
			$a_{u_v}$              & 6.806 $\times 10^{-3}$ &  &  &  &  &  &  & & \\
			\hline
			$b_{u_v}$              & 1.288 $\times 10^{-4}$  &  1.960 $\times 10^{-4}$ &  &  &  &  &  & & \\
			\hline
			$a_{d_v}$              & -3.128 $\times 10^{-5}$ & -4.824 $\times 10^{-5}$ & 8.880 $\times 10^{-4}$ &  &  &  &  &&  \\
			\hline
			$b_{d_v}$              &-6.200 $\times 10^{-4}$  & -2.449 $\times 10^{-4}$ &   5.015 $\times 10^{-4}$ &  1.254E-2 &  &  &  & & \\
			\hline
			$\eta_{\bar{q}}$       & 7.475 $\times 10^{-6}$ &  2.480 $\times 10^{-6}$  & 3.580 $\times 10^{-7}$   &  5.576 $\times 10^{-6}$  &3.294 $\times 10^{-3}$     &  &  & & \\
			\hline
			$a_{\bar{q}}$    	   & -2.266 $\times 10^{-4}$ & -6.417 $\times 10^{-5}$ & -1.587 $\times 10^{-4}$ &  5.096 $\times 10^{-3}$  & -1.716 $\times 10^{-6}$   & 0.9158  &  & & \\
			\hline
			$\eta_g$               &6.024 $\times 10^{-2}$ & 0.1448    & -1.891 $\times 10^{-2}$ & -0.8584    &  4.692 $\times 10^{-3}$  & -0.1003  & 8.065 $\times 10^{-4}$ & & \\
			\hline
			$a_g$                  & 2.040 $\times 10^{-3}$ & 5.0283 $\times 10^{-3}$  & -6.372 $\times 10^{-4}$ &-2.602 $\times 10^{-2}$   & 1.471 $\times 10^{-4}$    &-3.088 $\times 10^{-3}$ & 1.685 $\times 10^{-2}$ &  7.072 $\times 10^{-3}$ & \\
			\hline
			$\alpha_{s}(Q_0^2)$    &-1.722 $\times 10^{-3}$ & -3.997 $\times 10^{-3}$  & 5.322 $\times 10^{-4}$  &2.323 $\times 10^{-2}$   & -1.219 $\times 10^{-4}$   & 3.202 $\times 10^{-3}$ &  -1.360 $\times 10^{-3}$ & -4.604 $\times 10^{-2}$ & 1.296 $\times 10^{-3}$   \\
			\hline \hline
		\end{tabular}
	}
	\caption[]{ The covariance matrix for the $8 + 1$ free parameters in the NLO fit. }
	\label{covmat-matNLO}
\end{table*}
%
%

The Hessian or covariance matrix elements for 9 free parameters of our NNLO and NLO analysis which are obtained by running the CERN program library {\tt MINUIT} are given in Table~\ref{covmat-matNNLO} and \ref{covmat-matNLO}.
The uncertainties of PDFs can be calculated using these covariance matrix elements based on the Hessian method which can been used as a general
statistical method for estimating errors.
The uncertainty of polarized PDFs $f(x, \zeta)$ with respect to the optimized parameters $\zeta$ is then calculated by
using Hessian matrices and assuming linear error propagation,
\begin{eqnarray}\label{hessian-delta-K2}
	[\delta f(x)]^2 =
	\Delta \chi^2 \sum_{i,j}  \left( \frac{\partial f(x, \zeta)} {\partial \zeta_i}
	\right)_{ \zeta = \hat{\zeta}}  H_{ij}^{-1}  \left( \frac{ \partial f(x, \zeta)}{\partial \zeta_j} \right)_{\zeta = \hat{\zeta}} \,, \nonumber \\
\end{eqnarray}
where the $H_{ij}$ are the elements of the Hessian matrix, $\zeta_i$ is the quantity referring to the parameters which exist in polarized PDFs and $\hat \zeta$ indicates the number of parameters which make an extremum value for the related derivative.  The polarized PDFs uncertainties $\delta \Delta f(x, Q^2)$ at higher $Q^2$ scale are calculated by the well-known DGLAP evolution kernel.
The Hessian method which is based on the covariance matrix diagonalization, provides an efficient and simple method for calculating the PDFs uncertainties~\cite{Khanpour:2016pph,AtashbarTehrani:2012xh,Pumplin:2001ct,Martin:2002aw,Arbabifar:2013tma,Monfared:2011xf}.
In this method, one can assume that the deviation in the global goodness-of-fit quantity, $\Delta \, \chi^2_{\mathrm global}$, is quadratic in the deviation of the parameters specifying the input parton distributions $\zeta_i$ from their values at the minimum $\zeta_i^{\mathrm 0}$. One can write,
\begin{equation}\label{delta-chi}
	\Delta \chi_{\mathrm global}^2 \equiv \chi^2 - \chi_0^2 = \sum_{i, j} H_{ij}(\zeta_i - \zeta_i^{\mathrm  0 })(\zeta_j - \zeta_j^{\mathrm 0 }) \,,
\end{equation}
By having a set of appropriate polarized PDFs fit parameters which minimize the global $\chi ^ 2$ function, $s^{\rm 0}$, and introducing polarized parton sets $s^{\pm}_k$, one can write
\begin{equation}\label{pi}
	\zeta_i (s_{k}^{\pm }) = \zeta_i(s^{\rm 0 }) \pm t\sqrt{\lambda_k} \, v_{ik} \,,
\end{equation}
where $\lambda_k$ is the $k^{\rm th}$ eigenvalue and $v_{ik}$ is a set of orthonormal eigenvectors.
The parameter $t$ is adjusted to make the required $ T^2 = \Delta \chi_{\mathrm global}^2$ which is the allowed deterioration in $\Delta \, \chi _{\mathrm global}^2$
quality for the error determination and $t = T$ is the ideal quadratic behavior.
To test the quadratic approximation of Eq.~(\ref{delta-chi}), we study the dependence
of $\Delta \, \chi _{\mathrm global}^2$ along some random samples of eigenvector directions.
The $\Delta \chi_{\mathrm global}^2$ treatment for some selected eigenvectors, $k ^ {\mathrm {th} }$, numbered $k$ = 3, 4, 6 and 7 for the presented polarized PDFs analysis are illustrated in Fig.~\ref{fig:chi2}.
The detailed discussions on error estimation via Hessian method and an investigation of the quadratic
behavior of $\Delta \chi _{\mathrm global}^2$ can be found in Refs.~\cite{Khanpour:2016pph,AtashbarTehrani:2012xh,Martin:2002aw,Monfared:2011xf,Arbabifar:2013tma,Pumplin:2001ct}.
Although technical details are described in mentioned references, we prefer to explain outline of the Hessian method because it is used in our analysis.

The results of our polarized PDFs determination and error estimations will be discussed in much more details in Section.~\ref{Results-of-PPDFs}.

%
%
\section{Results of the NNLO polarized PDFs fits}\label{Results-of-PPDFs}
As stated in the Introduction, we intend to study the NNLO polarized PDFs consequently almost all polarized parton distributions in this work are presented in the NNLO order of QCD.
In this section we will present and discuss the results of our NNLO QCD analysis to the available world data on polarized inclusive DIS including the up-to-date data from {\tt COMPASS16} proton data~\cite{Adolph:2015saz}. Final parameter values for our NNLO and NLO QCD fits and their statistical errors in the $\overline{{\rm MS}}$--scheme at the input scale $Q_0^2$ = 4 GeV$^2$ are presented in Tables.~\ref{table:fitNNLO} and \ref{table:fitNLO}, respectively. Note that only the experimental errors (including systematic and statistical) are taken into account in this calculations. As seen from the Tables.~(\ref{table:fitNNLO}~ \ref{table:fitNLO}), the values of the parameters connected to the polarized PDFs are well determined. The quality of the fit can be judged from the obtained parameters and the $\chi^2$ values. There is a strong relationship between the input polarized PDFs parameterization and the uncertainties
which will be obtained. The parameterization for the input polarized PDFs in our analysis were presented in Section.~\ref{global-PPDFs}, specifically in Eq.~\ref{eq:parametrizations}. The free PDF parameters listed there allow a very large degree of flexibility.
The first moments of the polarized valence quark densities introduced in Eq.~\ref{eq:firstmoments} can be obtain by the constraints presented in Eq.~\ref{eq:constrain} which lead to the values of $\eta_{u_v} = 0.928 \pm 0.014$ and $\eta_{d_v} = -0.342 \pm 0.018$ for the $\Delta u_v$ and $\Delta d_v$ polarized valence distributions, respectively.

\begin{figure}[htb]
\vspace*{0.5cm}
\includegraphics[clip,width=0.5\textwidth]{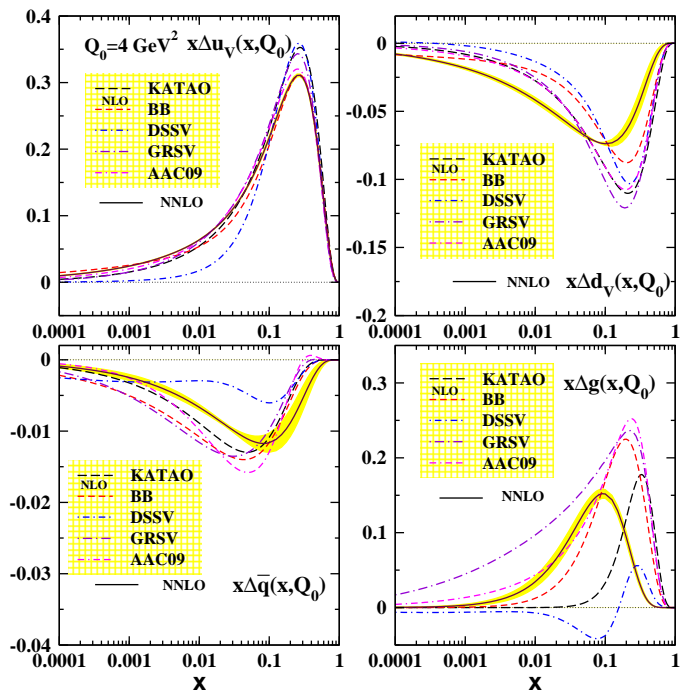}
\begin{center}
\caption{{\small (color online) Our results for the polarized parton distribution at Q$_0^2=$ 4 GeV$^2$
as a function of $x$ in NNLO approximation plotted as a solid curve. Also shown are
the results of {\tt KATAO} (long dashed)~\cite{Khorramian:2010qa}, {\tt BB} (dashed)~\cite{Blumlein:2010rn}, {\tt DSSV}~(dashed-dotted)
\cite{deFlorian:2008mr}, {\tt GRSV} (long dashed-dotted)~\cite{Gluck:2000dy},
and {\tt AAC09} (dashed-dashed-dotted)~\cite{Hirai:2008aj} in NLO approximation. \label{fig:PPDFsQ0}}}
\end{center}
\end{figure}

\global \long \def \fstrut{\rule{0pt}{12pt}}
\begin{table}
	\caption{\small  The paratemters of the NNLO input polarized PDFs at Q$_0^2$ = 4 GeV$^2$ obtained from the best fit to the available DIS data presented in Table.~\ref{tab:DISdata}. The details of the $\chi^2$ analysis and the constraints applied to control the parameters are contained in the text. \label{table:fitNNLO} }
	\begin{tabular}{>{\centering}p{0.3in}>{\centering}p{0.3in}c>{\centering}p{0.3in}>{\centering}p{0.3in}c}
		\hline\hline
		& $\eta_{u_v}$  & $~0.928 \, ({\rm fixed})~$  &  & $\eta_{\bar{q}}$  & $-0.04998 \pm 0.0497$   \\
		$\Delta u_v$  & $a_{u_v}$  &           $0.3915 \pm 0.0279$  & $\Delta \bar{q}$  & $a_{\bar{q}}$  & $0.4469 \pm 0.7992$  \\
		& $b_{u_v}$  & $3.1513 \pm 0.070$  &  & $b_{\bar{q}}$  & $4.954\, ({\rm fixed}) $   \\
		& $c_{u_v}$  & $10.675\, ({\rm fixed})$  &             & $c_{\bar{q}}$  & $0$       \\
		\hline
		& $\eta_{d_v}$  & $-0.342 \, ({\rm fixed})$  &  & $\eta_g$  & $0.3783 \pm 0.026$    \\
		$\Delta d_v$  & $a_{d_v}$  & $0.3677 \pm 0.022$  & $\Delta g$  & $a_{g}$  & $1.073 \pm 0.0903$    \\
		& $b_{d_v}$  & $4.923 \pm 0.563$  &  & $b_g$  & $10.705\, ({\rm fixed})$    \\
		& $c_{d_v}$  & $2.4107\, ({\rm fixed})$  &  & $c_g$  & $0$      \\
		\hline\hline
		\multicolumn{6}{c}{\fstrut $\alpha_s(Q_0^2) =\  0.275 \pm 0.024$ }     \\
		\hline\hline
		\multicolumn{6}{c}{\fstrut $\chi^{2}/{\rm dof}\ =\ 401.924/456\ =\ 0.881$ }   \\
		\hline\hline
	\end{tabular}
\end{table}
\global \long \def \fstrut{\rule{0pt}{12pt}}
\begin{table}
	\caption{\small  The paratemters of the NLO input polarized PDFs at Q$_0^2$ = 4 GeV$^2$ obtained from the best fit to the available DIS data presented in Table.~\ref{tab:DISdata}.\label{table:fitNLO} }
	\begin{tabular}{>{\centering}p{0.3in}>{\centering}p{0.3in}c>{\centering}p{0.3in}>{\centering}p{0.3in}c}
		\hline\hline
		& $\eta_{u_v}$  & $~0.928 \, ({\rm fixed})~$  &  & $\eta_{\bar{q}}$  & $-0.03224 \pm  0.0574$   \\
		$\Delta u_v$  & $a_{u_v}$  &           $ 0.230 \pm 0.0825$  & $\Delta \bar{q}$  & $a_{\bar{q}}$  & $ 0.5966 \pm 0.957$  \\
		& $b_{u_v}$  & $  2.6884 \pm  0.014$  &  & $b_{\bar{q}}$  & $7.661\, ({\rm fixed}) $   \\
		& $c_{u_v}$  & $ 21.10\, ({\rm fixed})$  &             & $c_{\bar{q}}$  & $0$    \\
		\hline
		& $\eta_{d_v}$  & $-0.342 \, ({\rm fixed})$  &  & $\eta_g$  & $0.6959 \pm 0.0284$    \\
		$\Delta d_v$  & $a_{d_v}$  & $0.3899 \pm 0.0298$  & $\Delta g$  & $a_{g}$  & $0.4575 \pm 0.0841$    \\
		& $b_{d_v}$  & $4.523 \pm  0.112$  &  & $b_g$  & $9.302\, ({\rm fixed})$    \\
		& $c_{d_v}$  & $3.899\, ({\rm fixed})$  &  & $c_g$  & $0$   \\
		\hline\hline
		\multicolumn{6}{c}{\fstrut $\alpha_s(Q_0^2) =\  0.2616 \pm 0.036 $}   \\
		\hline\hline
		\multicolumn{6}{c}{\fstrut $\chi^{2}/{\rm dof}\ =\ 410.856/456\ =\ 0.901$}   \\
		\hline\hline
	\end{tabular}
\end{table}
The extracted NNLO polarized PDFs are plotted in Fig.~\ref{fig:PPDFsQ0} for $x \Delta u_v$, $x \Delta d_v$,  $x \Delta \bar{q}$ and $x \Delta g$ distributions. The polarized PDFs are compared to those obtained at NLO analysis of {\tt KATAO} (long dashed)~\cite{Khorramian:2010qa}, {\tt BB} (dashed)~\cite{Blumlein:2010rn}, {\tt DSSV} (dashed-dotted)~\cite{deFlorian:2008mr}, {\tt GRSV}~(long dashed-dotted)~\cite{Gluck:2000dy} and {\tt AAC09} (dashed-dashed-dotted)~\cite{Hirai:2008aj}.
Examining the $x \Delta u_v$ and $x \Delta d_v$ polarized valence distributions, we see that most of the fits are in good agreements. However, our result for polarized valance distribution $x \Delta d_v$ is slightly smaller than others. For the  $x \Delta \bar{q}$ distribution, all of the curve except {\tt DSSV} are compatible.
Let us consider the plot for the polarized gluon distribution;
Due to the lack of experimental informations, the prediction for the small-$x$ behavior of the polarized gluon distribution $x \Delta g$, obtained form the different global analyses are largely uncertain. As the plots clearly show, the {\tt DSSV} result for the gluon distribution $x \Delta g$ has a sign change in the region of $x \sim 0.1$, while the other fits are positive. The {\tt DSSV} family polarized PDFs sets ({\tt DSSV}~\cite{deFlorian:2008mr,deFlorian:2009vb} and {\tt DSSV+/{\tt DSSV}++}~~\cite{deFlorian:2011ia,deFlorian:2014yva} ) include some of the non-DIS data of Table~\ref{tab:DISdata} such as SIDIS data, inclusive jet and hadron production measurements from polarized proton-proton measurements at RHIC collider. The plots also show that {\tt KATAO} for the gluon distribution approach to zero more quickly than the other results. The obtained NNLO polarized gluon distribution is slightly smaller as compared to the NLO analysis of {\tt BB}, {\tt GRSV} and {\tt AAC09}, and is positive for wide range of $x$; $x \gtrsim 0.001$. 

\begin{figure}[htb]
	\vspace*{0.5cm}
	\includegraphics[clip,width=0.5\textwidth]{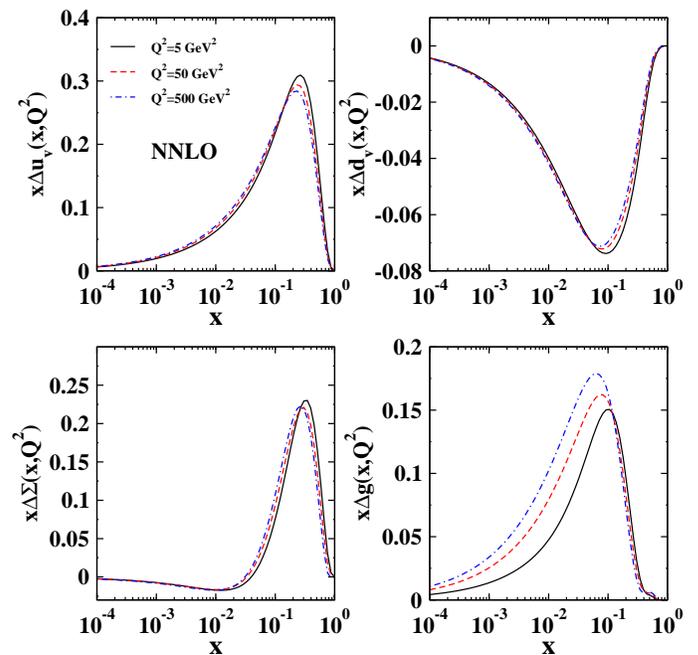}
	\begin{center}
		\caption{{\small (color online) The polarized parton distribution as function of $x$ and for different values of Q$^2$ = 5, 50, 500 GeV$^2$. \label{fig:partondiffQ}}}
	\end{center}
\end{figure}

In Fig.~\ref{fig:partondiffQ}, we plot the polarized parton distributions as a function of $x$ and for different values of Q$^2$ = 5, 50, 500 GeV$^2$. The plot predict an increase of gluon distribution in the kinematic region of $10^{-4}<x<10^{-1}$ by increasing the Q$^2$ values.

\begin{figure}[htb]
	\vspace*{0.5cm}
	\includegraphics[clip,width=0.5\textwidth]{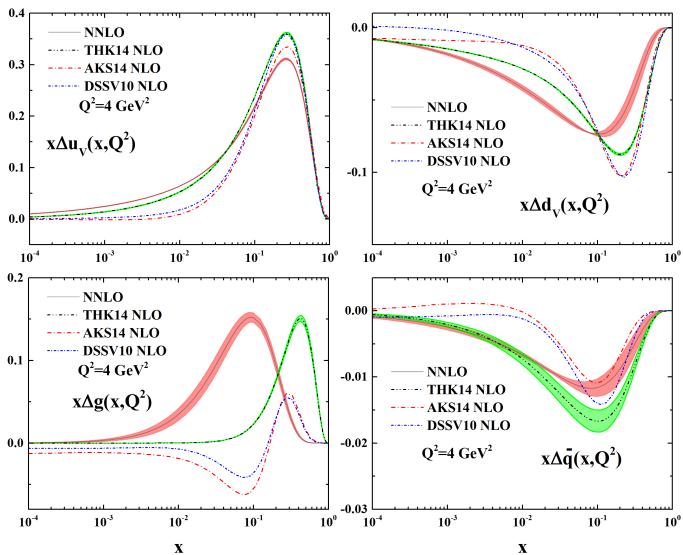}
	\begin{center}
		\caption{{\small (color online) Our polarized PDFs at Q$_0^2$ = 4 GeV$^2$
				as a function of $x$ in NNLO approximation plotted as a solid curve. Also shown are
				the most recent results from {\tt AKS14}~\cite{Arbabifar:2013tma}, {\tt THK14}~\cite{Monfared:2014nta} and {\tt DSSV10}~\cite{deFlorian:2009vb}. \label{fig:partonQ0tmcarbabifar}}}
	\end{center}
\end{figure}

In Fig.~\ref{fig:partonQ0tmcarbabifar}, we present our polarized parton distributions at Q$_0^2$ = 4 GeV$^2$ as a function of $x$ in NNLO approximation plotted as a solid curve. Also shown are the most recent results from {\tt AKS14}~\cite{Arbabifar:2013tma}, {\tt THK14}~\cite{Monfared:2014nta} and the {\tt DSSV} family polarized PDFs set~\cite{deFlorian:2009vb}. We also illustrate the uncertainties corresponding to the mentioned analysis.   Comparing to other results, one finds that the uncertainty band for our NNLO polarized sea distribution $x \Delta \bar{q}$ for low value of $x \lesssim 10^{-2}$ has become slightly narrower than {\tt THK14}. 
From the plots, we see that $x \Delta u_v$ and $x \Delta d_v$ are reasonably in agreement. The {\tt AKS14} and {\tt DSSV10} polarized valence distributions $x \Delta u_v$ slightly approach to zero more quickly than others.
For the polarized gluon distribution all of the fits treat differently. The ambiguity in gluon distributions may due to the different theoretical input and also the different data included in the QCD analysis. The {\tt AKS14} and {\tt DSSV10} polarized gluon distributions $x \Delta g$ have a sign change at $x \approx 0.2$. We had this behavior for {\tt DSSV08}~\cite{deFlorian:2008mr} polarized gluon distributions presented in Fig.~\ref{fig:PPDFsQ0}.
The striking feature of our NNLO polarized gluon distribution is its positivity throughout and clearly away from zero in the regime $x \gtrsim 0.0001$ predominantly probed by the RHIC and {\tt COMPASS} data. RHIC data mainly probe the region 0.05 $\lesssim x \lesssim$ 0.2, but the recently published data from {\tt COMPASS16}~\cite{Adolph:2015saz} which can cover the range of $0.0035 \lesssim x \lesssim 0.575$ can constrain $x \Delta g$ better down to somewhat lower values of $x \lesssim 0.02$, as we expected form this analysis. Overall, due to lack of enough data for low value of $x$, the constraints on $x \Delta g$ in, say, the regime 0.001 $\lesssim x \lesssim$ 0.05 are still much weaker than those in the region of $x > 0.2$.

\begin{figure}[htb]
	\vspace*{0.5cm}
	\includegraphics[clip,width=0.5\textwidth]{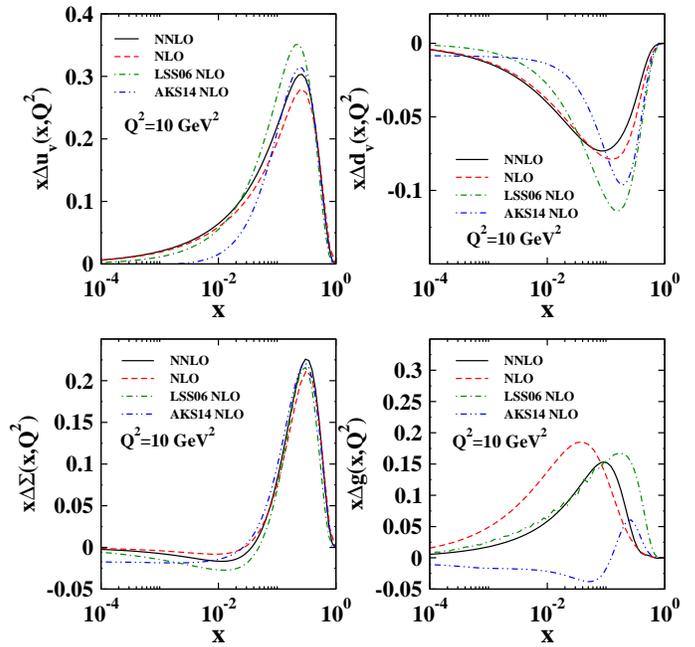}
	\begin{center}
		\caption{{\small (color online) The polarized parton distribution as a
				function of $x$ at Q$^2$=10 GeV$^2$ in NNLO approximation plotted as a solid curve. Also shown are
				the most recent results from {\tt AKS14}~\cite{Arbabifar:2013tma} and {\tt LSS06}~\cite{Leader:2006xc} analysis. Our revisited NLO analysis also shown as well. \label{fig:partonQ10}}}
	\end{center}
\end{figure}

In order to have a detailed comparison, we also plotted the obtained NNLO polarized PDFs as a function of $x$ at Q$^2$=10 GeV$^2$ which is presented in Fig.~\ref{fig:partonQ10}. The recent results from {\tt AKS14}~\cite{Arbabifar:2013tma} and {\tt LSS06}~\cite{Leader:2006xc} analysis also shown. 
Due to recent high precision measurements we also revisit our next-to-leading order QCD analysis of polarized PDFs. In this plot, we also illustrate our revisited NLO polarized PDFs results which have been extracted using the data presented in Table.~\ref{tab:DISdata}. The plot shows both of our NLO and NNLO polarized gluon distribution are positive throughout the $x$ range.

\begin{figure}[htb]
	\vspace*{0.5cm}
	\includegraphics[clip,width=0.5\textwidth]{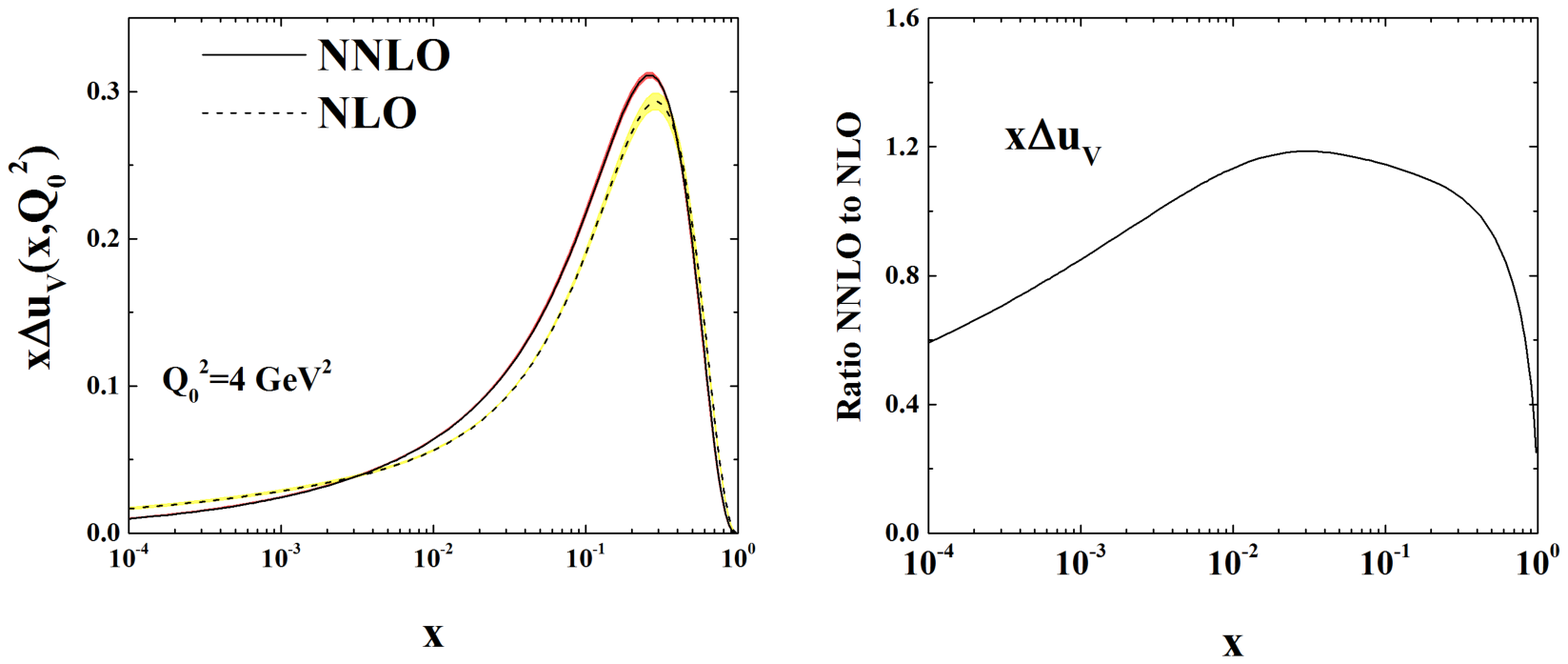}
	\includegraphics[clip,width=0.5\textwidth]{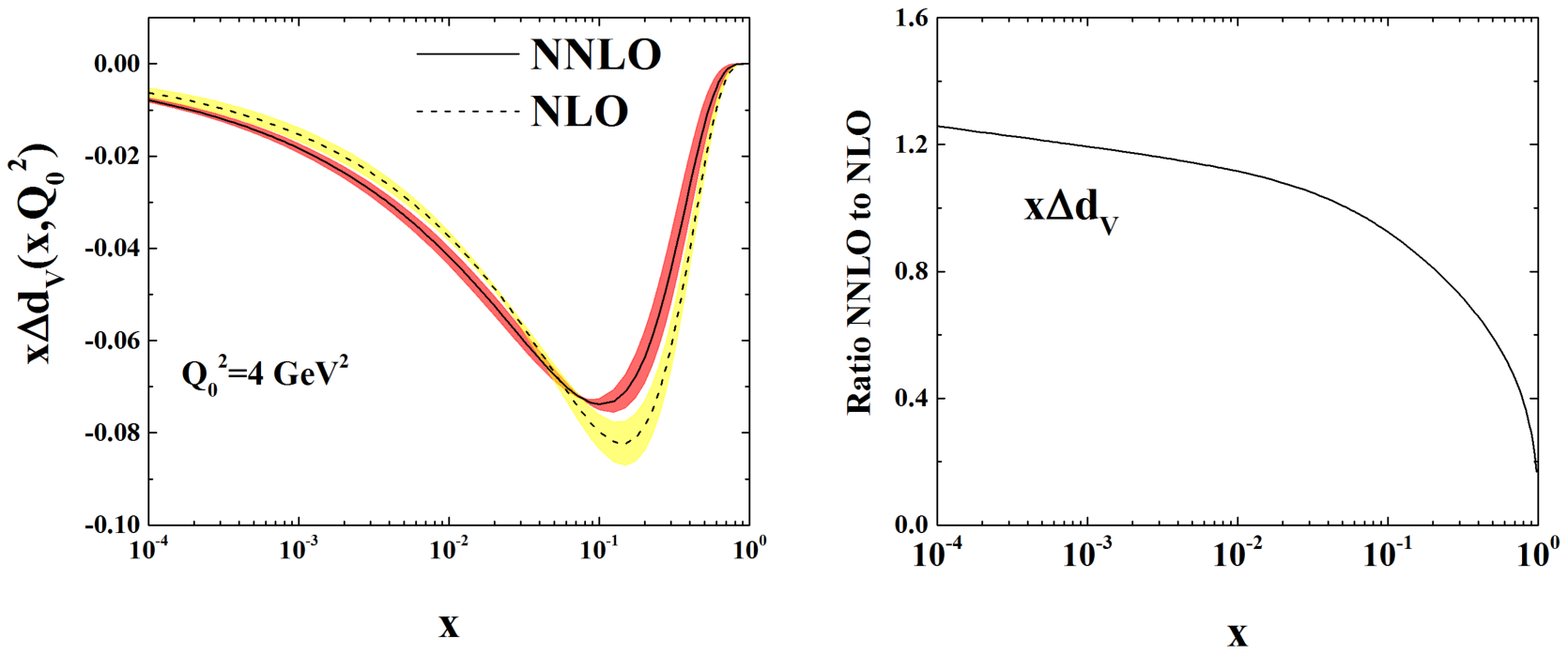}	
	\begin{center}
		\caption{{\small (color online) Comparison of the NNLO polarized up and down valence distributions (together with their uncertainties) with the
				NLO distribution at Q$_0^2$ = 4 GeV$^2$ (left), and the corresponding ratios for both the up and the down (right). All uncertainty
				bands represent a 90\% C.L. limit. \label{fig:ratio1}}}
	\end{center}
\end{figure}

\begin{figure}[htb]
	\vspace*{0.5cm}
	\includegraphics[clip,width=0.5\textwidth]{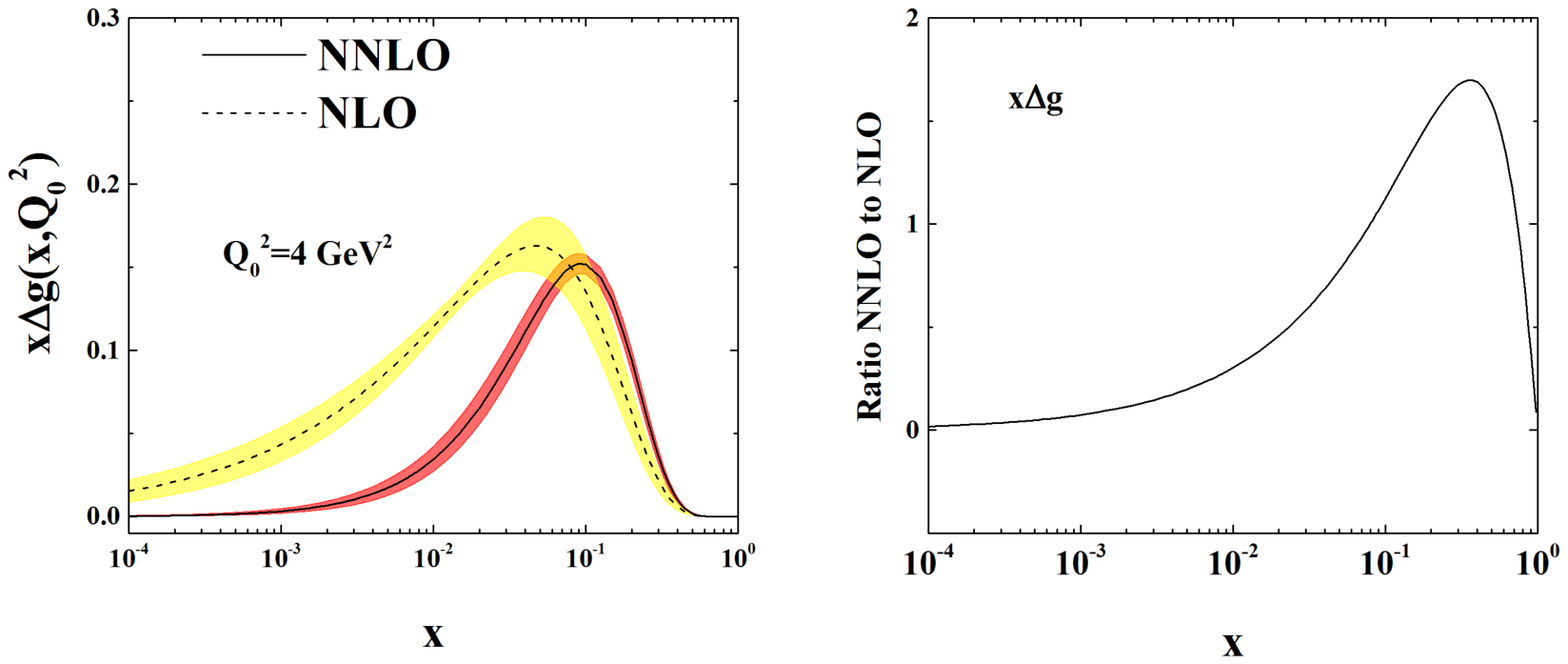}
	\includegraphics[clip,width=0.5\textwidth]{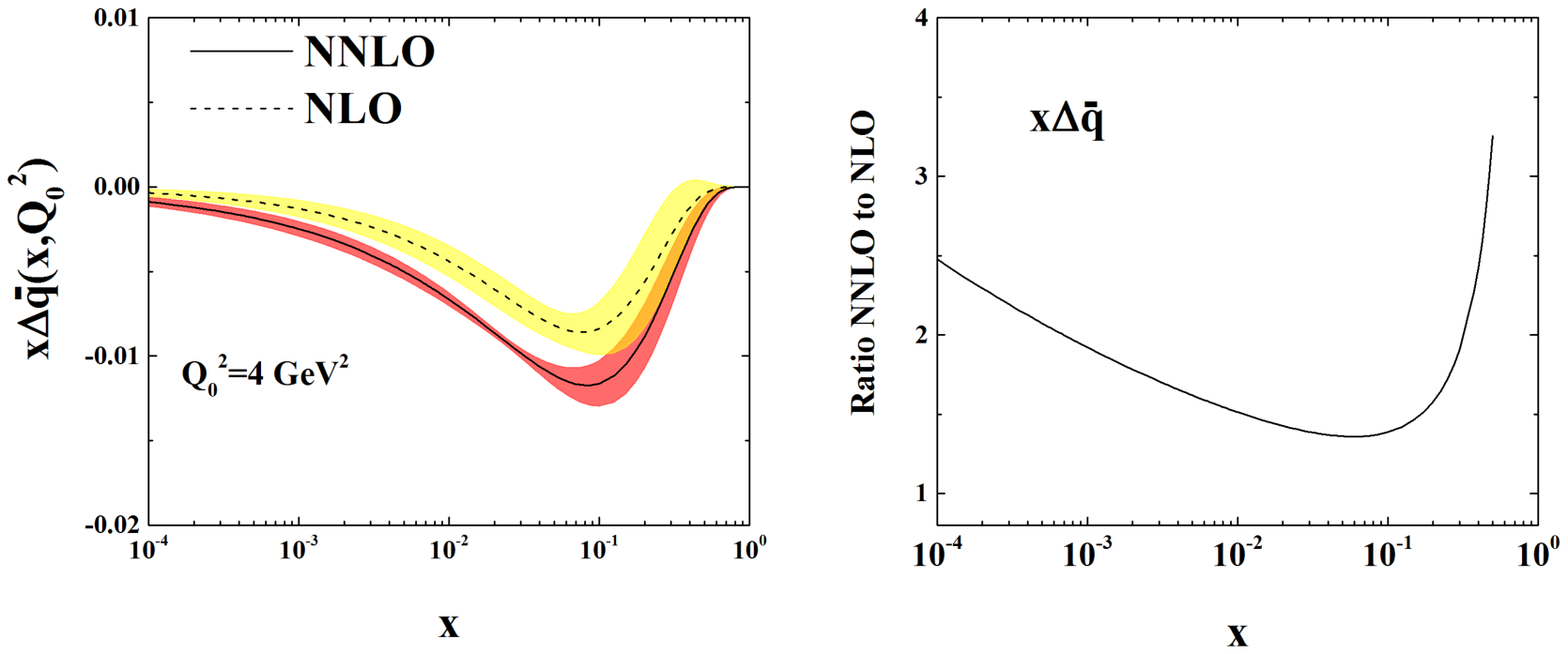}	
	\begin{center}
		\caption{{\small (color online) Comparison of the NNLO polarized sea and gluon distributions (together with their uncertainties) with the
				NLO distribution at Q$_0^2$ = 4 GeV$^2$ (left), and the corresponding ratios for both the sea and the gluon (right). All uncertainty
				bands represent a 90\% C.L. limit. \label{fig:ratio2}}}
	\end{center}
\end{figure}

What makes this analysis different from others are using the higher order QCD corrections and the inclusion of more precise data especially the most recent low-$x$ data from {\tt COMPASS16} experiments. In order to get an idea of the impact of higher order corrections and to examine the effect of the change in the NNLO polarized PDFs, we compare our NLO and NNLO analyses which have been extracted from the same data set using exactly the same functional forms for polarized distributions and the same assumptions. In this respect, we plot the polarized valence distributions in Fig.~\ref{fig:ratio1} and polarized sea and gluon distributions in Fig.~\ref{fig:ratio2}, respectively. The uncertainty bands for NLO and NNLO at 90\% C.L. limit which are obtained using the same approach for the input parameterization and error propagation, are also shown as well. In order to illustrate the significance of the size of the differences, we plot the ratios
of NNLO polarized PDFs to the corresponding NLO one in the right side of these Figures. The higher order QCD corrections lead to a significant change in the polarized gluon and sea distributions and in the obtained uncertainties. Moreover, there is most improvement in the description of the low-$x$ polarized distributions.  It is worth mentioning that the uncertainties of the polarized gluon PDFs at low value of $x$ still remain large compared to the currently probed region.


In order to examine the effect of PDF parameterisation on the obtained PDF uncertainty, especially for the polarised gluon
distribution, we release a few more of the parameters such as the $b_g$ and $c_g$, 
\begin{equation}
x \, \Delta g (x, Q_0^2) = \eta_g x^{a_g} (1-x)^{b_g} (1 + c_g x) \,.
\end{equation}
The result shows that despite of increasing the uncertainties, the shape of the polarized gluon distribution will not changed. In the present work, we only used the usual parametrization for the gluon distribution to have a detailed comparison with the results from {\tt THK14}~\cite{Monfared:2014nta} and {\tt KATAO}~\cite{Khorramian:2010qa}.
The difficulties in constraining the polarized gluon distribution are clearly revealed through the spread of $x \Delta g$ from various global PDFs parametrizations illustrated in Figs.~\ref{fig:PPDFsQ0} and \ref{fig:partonQ0tmcarbabifar}. These plot clearly show that depending on the global PDFs parametrizations, the method of PDFs uncertainty estimation and the data sets included in the fits, the shapes and magnitudes from the gluon PDFs including its uncertainty are generally different.  
In most of the fits the $x \Delta g$ PDFs is positive at large value of $x$ with a sign change at smaller value of $x$ for {\tt THK14} and the {\tt DSSV}. 
In both our NLO and NNLO analysis, we obtained a positive $x \Delta g$ PDFs which is clearly away from zero in the regime $x \gtrsim 0.0001$.          

The best-fit values of the first moments of $g_1$ structure function can be obtained using the analyzed polarized PDFs. One can determine the first moment as,
\begin{equation}
\Gamma_1^p (Q^2)  \equiv \int _0^1 dx \, g_1^p (x, Q^2) \,.
\end{equation}
The corresponding results for the first moments using the extracted polarized PDFs are presented in Table.~\ref{table1:firstMomQ} for selected values of Q$^2$.
\begin{table}[htb]
\caption{\small The best-fit values of first moments for the polarized PDFs, $\Delta u_v$, $\Delta d_v$, $\Delta \overline{q}$, $\Delta g$ and polarized
structure functions $\Gamma_1^p$, $\Gamma_1^n$ and $\Gamma_1^d$
in NNLO approximations in the $\overline{{\rm MS}}$--scheme for some different values of Q$^2$. \label{table1:firstMomQ}}	
	\begin{tabular}{ccccc}
		\hline
		\hline
		Q$^2$  & 2 GeV$^2$  & 5 GeV$^2$  & 10 GeV$^2$  & 50 GeV$^2$  \\
		\hline
		\hline
		$\Delta u_v$  &  0.92644  & 0.92589  &  0.92562  &  0.92508      \\
		$\Delta d_v$  &  -0.34116  &  -0.34096  &  -0.34086  & -0.34066    \\
		$\Delta \Sigma$  &  0.285276  &  0.285105  & 0.285019  &  0.28485    \\
		$\Delta g$    &  0.33012  & 0.39138  &  0.426678  & 0.50931     \\ \hline
		$\Gamma_1^p$  &  0.12187  & 0.13229  & 0.13673  &  0.14393     \\
		$\Gamma_1^n$  & -0.05332  & -0.05441  & -0.05492  &  -0.05582   \\
		$\Gamma_1^d$  & 0.031706  & 0.036019  &   0.037840  & 0.040752   \\
		\hline
	\end{tabular}
\end{table}

The numerical results for the polarized structure functions $\Gamma_1^p$, $\Gamma_1^n$ and $\Gamma_1^d$
in NNLO approximations are compared with the corresponding data from recently published {\tt COMPASS16} results in Table.~\ref{table2:firstMomQ}.
The table clearly shows our results describe the experimental measurements well.
\begin{table}[htb]
\caption{ \small First moments of $g_1$ at Q$^2$ = 3 GeV$^2$ presented by {\tt COMPASS16}~\cite{Adolph:2015saz} experiment. Our NNLO theory predictions also shown as well.~\label{table2:firstMomQ} }	
	\begin{tabular}{cccc}
		\hline\hline
	                 &              {\tt COMPASS16}                    &     NNLO (MODEL)          &   \tabularnewline
		$\Gamma^p$   &         0.139 $\pm$ 0.003 $\pm$ 0.009           &     0.12742               &   \tabularnewline
		$\Gamma^n$   &        -0.041 $\pm$ 0.006 $\pm$ 0.011           &    -0.05389               &   \tabularnewline
		$\Gamma^{\rm NS}$&     0.181 $\pm$ 0.008 $\pm$ 0.014           &     0.18131               &   \tabularnewline
		\hline
	\end{tabular}
\end{table}

\begin{table}[htb]
	\caption{The $\alpha_s(M_Z^2)$ values in comparison with the results obtained by other QCD analyses of inclusive deep-inelastic scattering
		data in NLO, NNLO and NNNLO approximations.\label{tab:as}}		
	\begin{tabular}{cccc}
		\hline
		$\alpha_{s}(M_{Z}^{2})$  & Order  & Reference  & Notes \tabularnewline
		\hline \hline
		$0.1169 \pm 0.0006$  & NLO  &   &                       This analysis            \\
		$0.1132_{-0.0095}^{+0.0056}$  & NLO  & \cite{Blumlein:2010rn}  &  {\tt BB10}     \\
		$0.1149 \pm 0.0015$  & NLO  & \cite{Khorramian:2010qa} & {\tt KATAO}                   \\
		$0.1141 \pm 0.0036$  & NLO  & \cite{AtashbarTehrani:2007odq}  &  {\tt TK}              \\
		$0.1180$           & NLO  & \cite{Owens:2012bv}  & {\tt CJ12}                          \\
		$0.1136 \pm 0.0012$  & NLO  & \cite{Monfared:2014nta}  &  {\tt THK14}                 \\
		$0.1142 \pm 0.0014$           & NLO  & \cite{Khanpour:2012tk}  & {\tt KKT12C}          \\
		$0.1150 \pm 0.0018$           & NLO  & \cite{Khanpour:2012tk}  & { \tt KKT12}           \\   \hline
		$0.1186 \pm 0.0005$  & NNLO  &  &                       This analysis            \\
		$0.1134_{-0.0021}^{+0.0019}$  & NNLO  & \cite{Blumlein:2006be}  &  {\tt BBG06}   \\
		$0.1131 \pm 0.0019$  & NNLO  & \cite{Khorramian:2008yh}  &  {\tt KT08}                 \\
		$0.1135 \pm 0.0014$  & NNLO  & \cite{ABKM}  & {\tt ABKM10-FFS}                                \\
		$0.1129 \pm 0.0014$  & NNLO  & \cite{ABKM}  & {\tt ABKM10-BSM}                            \\
		$0.1124 \pm 0.0020$  & NNLO  & \cite{JR}  & {\tt GRS} dynamic approach                     \\
		$0.1158 \pm 0.0035$  & NNLO  & \cite{JR}  & {\tt GRS} standard approach                    \\
		$0.1171 \pm 0.0014$  & NNLO  & \cite{MSTW}  &  {\tt  MSTW08 }                          \\
		$0.1145 \pm 0.0042$  & NNLO  & \cite{H1ZEUS}  &  H1 and ZEUS                     \\
		$0.1177 \pm 0.0013$  & NNLO  & \cite{d'Enterria:2015toz}  & Preliminary          \\ \hline
		$0.1139 \pm 0.0020$  & NNNLO  & \cite{Khorramian:2009xz}  & {\tt KKT10}                \\
		$0.1141_{-0.0022}^{+0.0020}$  & NNNLO  & \cite{Blumlein:2006be}  &   {\tt  BBG06}        \\ \hline
		$0.1185 \pm 0.0006$  & ---   & \cite{Agashe:2014kda}  & World Average            \\
		\hline \hline
	\end{tabular}
\end{table}

In order to check the accuracy of the extracted polarized parton distribution functions, we present the recent results for the running coupling constant in Table.~\ref{tab:as}. The results obtained by available QCD analysis of inclusive deep-inelastic scattering data in NLO, NNLO and NNNLO approximations including the current world average of $\alpha_{s}(M_Z^2) = 0.1185 \pm 0.0006$~\cite{Agashe:2014kda} are also presented.
Our results for the running coupling constant, $\alpha_s(M_Z^2)$, also shown as well. We obtained the following value for the strong coupling constant at Z boson mass scale at NNLO approximation,
\begin{equation}
\alpha_s(M_Z^2) = 0.1186 \pm 0.0005~.\label{eq:alphaMZ}
\end{equation}
The higher order QCD correction leads to a larger value of the QCD coupling at NNLO.
To close this section, we note that using simple forms of parametrization and enormous amount of constraining data, the NNLO distributions lead to a considerable decrease in the polarized PDFs uncertainties. Detailed comparisons to the various NLO sets and with the data, will be made in the next Section.

\begin{figure*}[htb]
	\vspace*{0.5cm}
	\includegraphics[clip,width=0.8\textwidth]{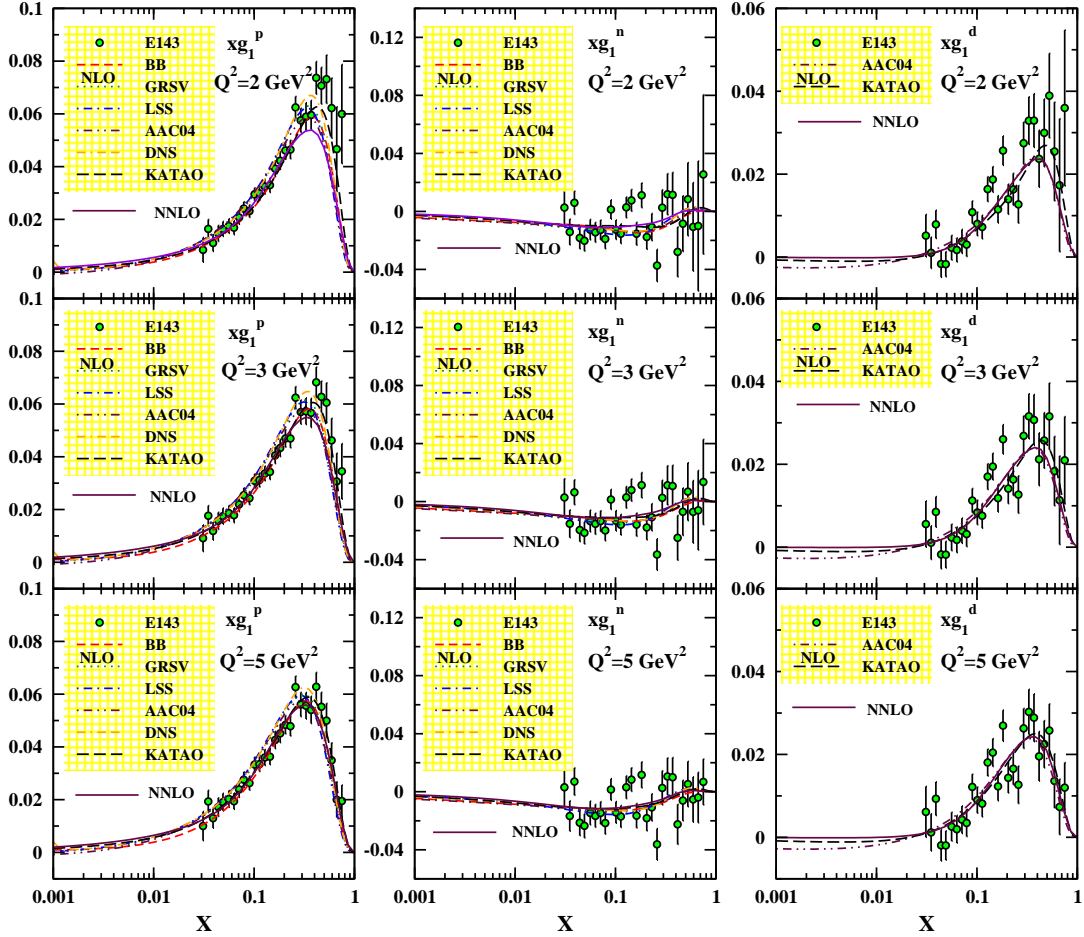}
	\begin{center}
		\caption{{\small (color online) The theory predictions for the polarized structure function $x g_1^{\rm p,n,d}$ as a function
				of Q$^2$ in intervals of $x$.  Our theory predictions are the solid curve in NNLO approximation. Also shown are
				the QCD NLO curves obtained by {\tt KATAO} (long dashed)~\cite{Khorramian:2010qa},  {\tt BB} (dashed)~\cite{Bluemlein:2002be},
				{\tt GRSV} (dotted)~\cite{Gluck:2000dy}, {\tt LSS} (dashed-dotted)~\cite{Leader:2005ci}, {\tt DNS} (dashed-dashed-dotted)~\cite{deFlorian:2005mw}
				and {\tt AAC04} (dashed-dotted-dotted)~\cite{Goto:1999by}. Data points are from
				the {\tt E143}~\cite{Abe:1998wq} experiments at SLAC. \label{fig:g1pnd}}}
	\end{center}
\end{figure*}
\begin{figure}[htb]
	\vspace*{0.5cm}
	\includegraphics[clip,width=0.5\textwidth]{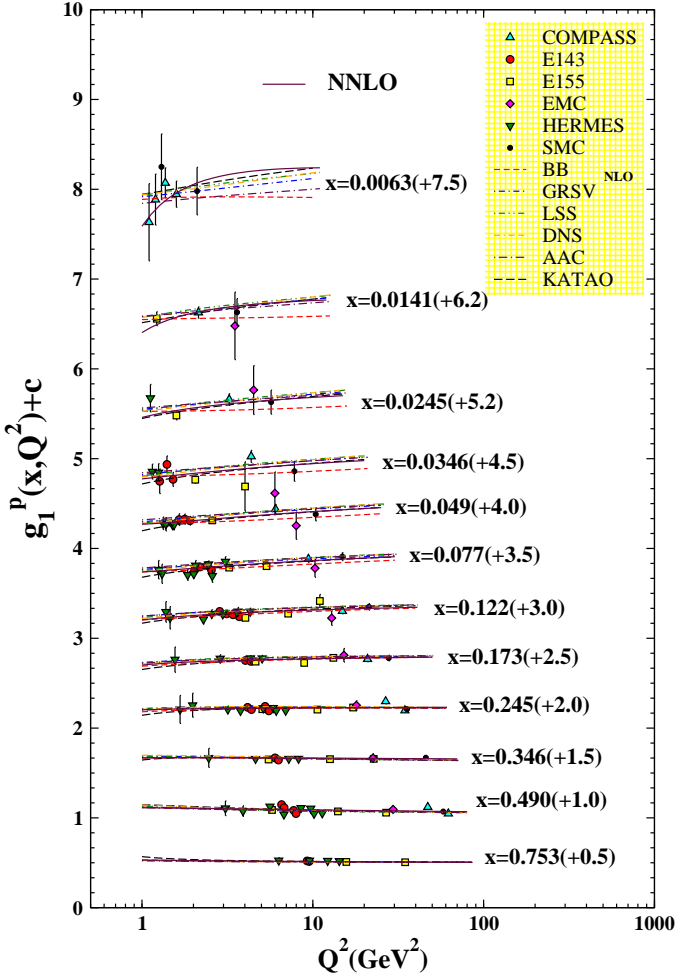}
	\begin{center}
		\caption{{\small (color online) The prediction for the polarized structure function $g_1^{p}$ as function
				of Q$^2$ in intervals of $x$. Our fit is the solid	curve in NNLO approximation. The error bars shown are the statistical
				and systematic uncertainties added in quadrature. The values of the shift $c$ are given in parentheses.
				Also shown are available experimental data and the results from {\tt KATAO} (long dashed)~\cite{Khorramian:2010qa}, {\tt BB} (dashed)~\cite{Bluemlein:2002be},
				{\tt GRSV} (dashed-dotted)~\cite{Gluck:2000dy}, {\tt LSS} (dashed-dotted-dotted)~\cite{Leader:2005ci}, {\tt DNS} (dashed-dashed-dotted)~\cite{deFlorian:2005mw}
				and {\tt AAC04} (long dashed-dotted)~\cite{Goto:1999by} in NLO approximation. \label{fig:g1p}}}
	\end{center}
\end{figure}

%
%
\section{Comparison with the data and different global analyses of polarized PDFs}\label{Comparison-with-the-data}
Throughout the above discussion we have presented our NLO and NNLO polarized PDFs including their uncertainties. 
In the following section, we will present a detailed comparison of our NNLO polarized PDFs with the data and other phenomenological models. 
In Fig.~\ref{fig:g1pnd}, the spin-dependent structure function of the proton, neutron and deuteron are displayed as a function of $x$ at Q$^2$ = 2, 3 and 5 GeV$^2$, respectively.
The solid curve represent to our theory predictions for $x g_1^{\rm p,n,d}$ at NNLO approximation. The results of those obtained at NLO from {\tt KATAO} (long dashed)~\cite{Khorramian:2010qa}, {\tt BB} (dashed)~\cite{Bluemlein:2002be}, {\tt GRSV} (dotted)~\cite{Gluck:2000dy}, {\tt LSS} (dashed-dotted)~\cite{Leader:2005ci}, {\tt DNS} (dashed-dashed-dotted)~\cite{deFlorian:2005mw}
and {\tt AAC04} (dashed-dotted-dotted)~\cite{Goto:1999by} also shown as well. Data points are from
the {\tt E143}~\cite{Abe:1998wq} experiments at SLAC. The good quality of the fits for the best-fit polarized structure functions are apparent from these plots.
The poor quality of current knowledge of the shape of polarized parton distributions and structure functions at $x \leq 0.01$ are a consequence of the limited kinematic coverage of polarized DIS data at small $x$.

The prediction for the polarized proton structure function $g_1^{\rm p}$ as function of Q$^2$ in intervals of $x$ is presented in Fig.~\ref{fig:g1p}. Our fit is the solid
curve in NNLO approximation. The error bars shown are the statistical and systematic uncertainties added in quadrature.
The values of the shift parameter $c$ are given in parentheses. Also shown are the results of {\tt KATAO} (long dashed)~\cite{Khorramian:2010qa}, {\tt BB} (dashed)~\cite{Bluemlein:2002be},
{\tt GRSV} (dashed-dotted)~\cite{Gluck:2000dy}, {\tt LSS} (dashed-dotted-dotted)~\cite{Leader:2005ci}, {\tt DNS} (dashed-dashed-dotted)~\cite{deFlorian:2005mw}
and {\tt AAC04} (long dashed-dotted)~\cite{Goto:1999by} in NLO approximation.

\begin{figure}[htb]
	\vspace*{0.5cm}
	\includegraphics[clip,width=0.5\textwidth]{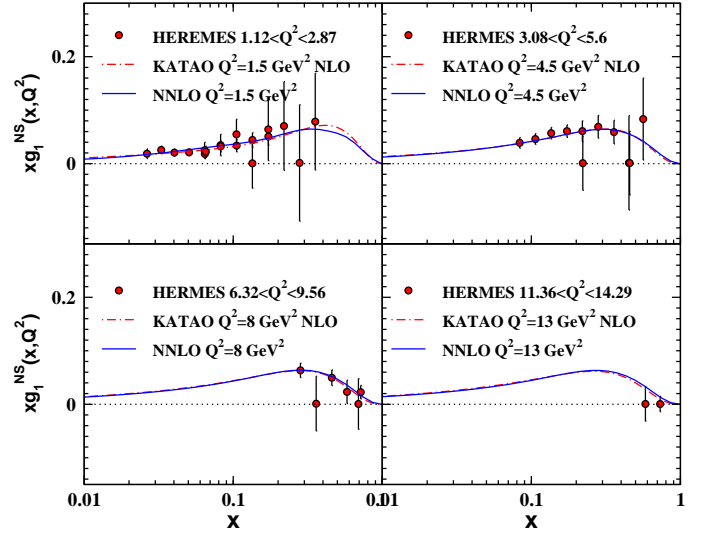}
	\begin{center}
	\caption{{\small  (color online) The prediction for the non-singlet polarized structure function $x g_1^{\rm NS}$
	as a function of $x$ in NNLO approximation in comparison with the NLO results of {\tt KATAO}~\cite{Khorramian:2010qa} model. Also shown is the up-to-date experimental data from {\tt HERMESS}~\cite{Airapetian:2006vy}. \label{fig:g1NS}}}
	\end{center}
\end{figure}

We are in positions to study the behavior of our NNLO polarized parton distributions functions in the regions of small and large momentum fractions.
Having investigated the neutron, proton and deuteron spin-dependent structure function, one can turn to the non-singlet spin-dependent structure function as,
\begin{eqnarray}\label{eq:g1NS}
x g_1^{\rm NS} (x, Q^2) &=&  x g_1^{\rm p} (x, Q^2) - x g_1^{\rm n} (x, Q^2)   \nonumber   \\
&=&  2 \left[ x g_1^{\rm p} (x, Q^2) - x g_1^{\rm N} (x, Q^2) \right]   \nonumber   \\
&=& 2 \left[ xg_1^{\rm p} (x, Q^2) - \frac{xg_1^{\rm d} (x, Q^2)}{1 - \frac{3}{2} w_D }  \right]  \,,
\end{eqnarray}
where $x g_1^{\rm N}$ is the nucleon structure function and $w_D = 0.05 \pm 0.01$ is the D-state wave probability for the deutron~\cite{Lacombe:1981eg,Machleidt:1987hj}.
The prediction for the non-singlet polarized structure functions $x g_1^{\rm NS}$ as function of $x$ in NNLO approximation are plotted in Fig.~\ref{fig:g1NS}. The NLO result from {\tt KATAO}~\cite{Khorramian:2010qa} shown for comparison. The plots correspond to the bin $1.12 < Q^2 < 2.87$, $3.08 < Q^2 < 5.60$,  $6.32 < Q^2 < 9.56$  and $11.36 < Q^2 < 14.29$,  respectively. The plots show that our results for the non-singlet polarized structure functions describe both the data and the results obtained by {\tt KATAO} analysis well.

\begin{figure}[htb]
	\vspace*{0.5cm}
	\includegraphics[clip,width=0.45\textwidth]{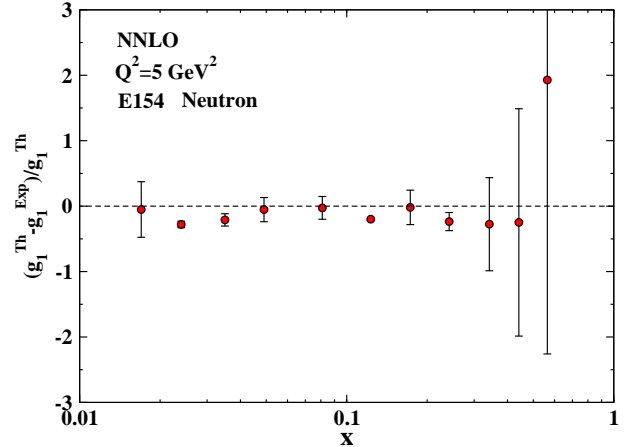}
	\begin{center}
		\caption{{\small (color online)  The ratios of $(g_1^{\rm Th} - g_1^{\rm Exp})/g_1^{\rm Th}$ are shown for comparison. The NNLO parametrization is used for the theoretical calculations at the Q$^2$ = 5 GeV$^2$ points of the experimental data for polarized neutron {\tt E154}~\cite{E154n}. \label{fig:g1nratio}}}
	\end{center}
\end{figure}

The prediction for the best-fit polarized neutron structure function $g_1^{\rm n}$ can also been obtained using analyzed polarized PDFs. In Fig.~\ref{fig:g1nratio}, we plot the ratios of $( g_1^{\rm Th} -  g_1^{\rm Exp})/ g_1^{\rm Th}$ where $ g_1^{\rm Exp}$ is the experimental value of the polarized neutron structure function and $ g_1^{\rm Th}$ is the corresponding theoretical values. Also shown are the most recent data from {\tt E154}~\cite{E154n} collaborations.

\begin{figure}[htb]
	\begin{center}
		\vspace{1cm}
		\resizebox{0.45\textwidth}{!}{\includegraphics{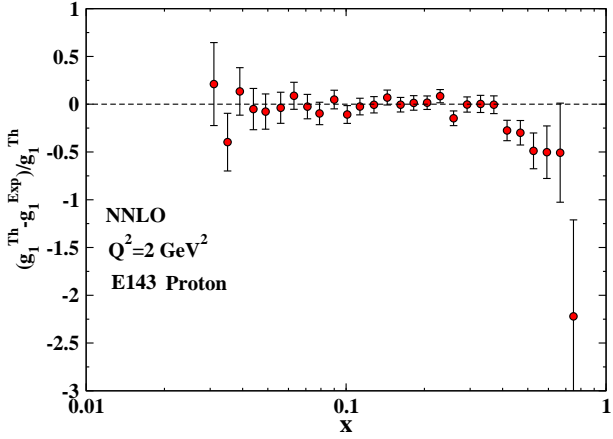}}
		\caption[]{ (Color online) Comparison with experimental data of {\tt E143}~\cite{Abe:1998wq}. The ratios of $(g_1^{\rm Th} - g_1^{\rm Exp})/g_1^{\rm Th}$ are shown for comparison. The NNLO parametrization is used for the theoretical calculations at the Q$^2$ = 2 GeV$^2$ points of the experimental data.}\label{fig:g1pratio}
	\end{center}
\end{figure}

A detailed comparison with the experimental data of the polarized deeply proton structure function for the analyzed polarized PDFs is also shown in Fig.~\ref{fig:g1pratio}.
The ratios of $(g_1^{\rm Th} - g_1^{\rm Exp})/g_1^{\rm Th}$ are shown for comparison. $g_1^{\rm Exp}$ is the experimental value and $g_1^{\rm Th}$ is the theoretical value of the polarized proton structure function. Also shown are the most recent data from {\tt E143}~\cite{Abe:1998wq} collaborations.

\begin{figure}[htb]
	\vspace*{0.5cm}
	\includegraphics[clip,width=0.40\textwidth]{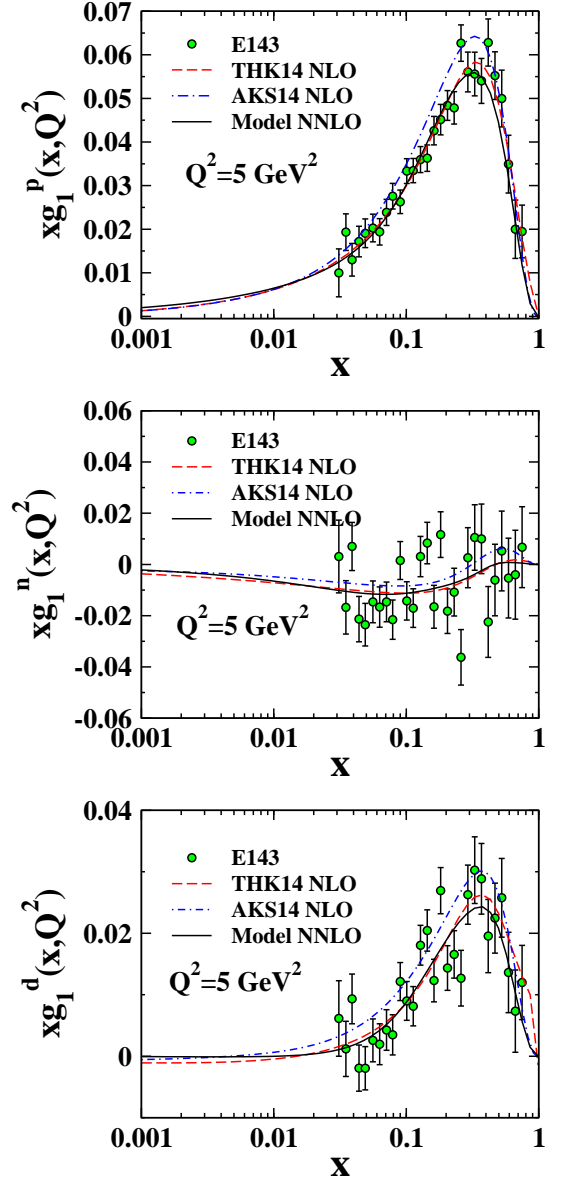}
	\begin{center}
\caption{{\small (color online) The prediction for the polarized nucleon structure function $x g_1^{\rm N} (x, Q^2) (\rm N = p, n, d)$ as a function
of $x$ at Q$^2$ = 5 GeV$^2$. Also shown are the most recent polarized global analysis from {\tt AKS14}~\cite{Arbabifar:2013tma} and {\tt THK14}~\cite{Monfared:2014nta}. Data points are from the {\tt E143}~\cite{Abe:1998wq} experiments at SLAC. \label{fig:xg1Q5tmc}}}
	\end{center}
\end{figure}

In Fig.\ref{fig:xg1Q5tmc}, we plot the polarized nucleon structure functions $x g_1^{\rm N} (x, Q^2) (\rm N = p, n, d)$ as a function of $x$ at Q$^2$ = 5 GeV$^2$.
Data points are from the {\tt E143}~\cite{Abe:1998wq} experiments at SLAC. For comparison, the most recent polarized global analysis from {\tt AKS14} and {\tt THK14} also shown as well. The {\tt THK14} analysis carried out a next-to-leading order QCD analysis to the polarized structure functions $g_1$ and $g_2$ and included the target mass corrections and higher twist effects in the analysis. {\tt AKS14} has presented a next-to-leading order QCD analysis of the polarized DIS and SIDIS data on the nucleon. They also considered the SU(2) and SU(3) symmetry breaking scenario. Examining the polarized proton structure function, $x g_1^p$, we see that our NNLO fits and {\tt THK14} are in satisfactory agreements. For $x g_1^n$ and  $x g_1^d$, we see our results slightly are smaller than {\tt THK14} and {\tt AKS14} for larger values of $x$. Overall the results show that all of the analysis perfectly describe the data well.

It is worth pointing out in this context that the plots presented above clearly show that the expected statistical accuracies are very good for all analyzed polarized DIS data. This suggests that a reasonable accurate determination of polarized structure function as well as polarized PDFs using Jacobi polynomials expansion approach is possible.

%
%
\section{Polarized PDFs at the dawn of the RHIC and LHC} \label{RHIC-LHC-era}

The past few years have witnessed tremendous progress in our understanding of the polarized DIS structure functions as well as polarized PDFs.
Recent {\tt PHENIX} measurements on the inclusive $\pi^0$ production in polarized proton-proton collisions at the Relativistic Heavy Ion Collider (RHIC) at center-of-mass energy of $\sqrt{s}$ = 510 GeV~\cite{Adare:2015ozj,Adare:2014hsq} as well as {\tt STAR} measurements at RHIC on inclusive jet production in polarized proton collisions at $\sqrt{s}$ = 200 GeV and double spin asymmetries from open charm muon production and leading and next-to-leading order gluon polarization determination in the nucleon at {\tt COMPASS}~\cite{Adolph:2012ca}, have led to significant improvement in the determination of the polarized gluon distributions especially at small value of $x$~\cite{deFlorian:2014yva}.
The new measurements from the {\tt PHENIX} experiments at RHIC on longitudinal single-spin asymmetries in W$^\pm$ and Z boson production collisions at center of mass energies of $\sqrt{s}$=500 and 510~GeV~\cite{Adare:2015gsd,Adamczyk:2014xyw} are also yielding better constraints on the polarization of sea quarks and anti-quarks.

\begin{figure}[htb]
	\vspace*{0.5cm}
	\includegraphics[clip,width=0.45\textwidth]{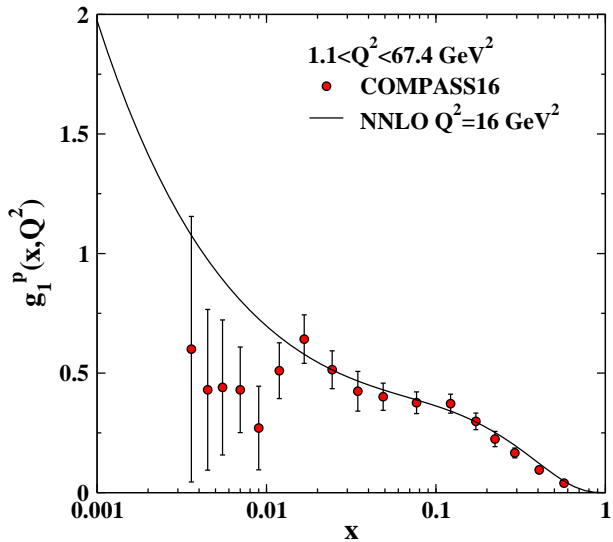}
	\begin{center}
		\caption{{\small (color online) The prediction for the polarized proton structure function $g_{1}^{\rm p}$ as function
				of $x$ and for mean value of Q$^2$ = 16 GeV$^2$. Also shown are the most recent data from {\tt COMPASS16} collaboration~\cite{Adolph:2015saz}. Note that the values of Q$^2$ for each data point are different. \label{fig:g1pCOMPASS}}}
	\end{center}
\end{figure}

Recently, {\tt  COMPASS} collaborations at CERN collected a large number of events of polarized inelastic scattering and presented their results for the proton longitudinal spin structure functions $g_1^p$ and the double spin asymmetry $A_1^p$~\cite{Adolph:2015saz}. These data collected at center-of-mass energy of $\sqrt{s}$ = 200 GeV which is the first data covers a very low values of $x$. The statistical precision of $g_1^p$ improved in the region $x \lesssim  0.02$. The mentioned data covers the range of $0.0035 \lesssim x \lesssim 0.575$.
The results of the QCD fits to the polarized proton structure function $x g_1^{\rm p}$ as a functions
of $x$ and for mean value of Q$^2$ = 16 GeV$^2$ is illustrated in Fig.~\ref{fig:g1pCOMPASS}. The solid curve represents our best-fit at NNLO approximation accuracy of perturbative QCD.
The data are taken from recent {\tt COMPASS16} measurements~\cite{Adolph:2015saz}.  Note that the values of Q$^2$ for each measured point are different.
The proposed high luminosity and high energy Electron-Ion collider (EIC) such as eRHIC~\cite{Aschenauer:2014cki,Miller:2015tjf} and EIC@HIAF~\cite{Yang:2013yeb} can probe a broad Q$^2$ < 1 GeV$^2$-range, where one can check the predicted behavior of g$_1^p$ at this region. The improved accuracy and the kinematic coverage of the future RHIC data form {\tt PHENIX} and  {\tt STAR} can lead to more precise polarized PDFs determination in common global QCD analysis of world data. In addition, the accuracy of present determination of polarized PDFs especially polarized gluon distribution still require a widening of the kinematic coverage at small $x$ which can be achieved at EIC~\cite{Accardi:2012qut}. Many new and important results reported by these experiments can change substantially our perception of the gluon helicity distribution.
For the future, there are more new and precise data to be included. This will lead us to produce fully updated NLO and NNLO polarized PDFs with uncertainties. However, until this major update can be finalized, the NNLO polarized PDFs outlined in this note will serve the only set currently available at NNLO.

%
%
\section{Summary and conclusions} \label{Summary}
In this comparative study, we wish to present for the first time a NNLO polarized PDFs analysis of the inclusive world data for $g_1^p$, $g_1^d$ and $g_1^n$ including the recently published {\tt COMPASS16} spin-dependent proton structure function.  To establish a meaningful baseline for estimating the impact of these DIS data, we also revisited our next-to-leading order QCD analysis. We have used the Jacobi polynomials expansion method to facilitate the analysis. Overall a very good description of the global inclusive polarized DIS data set has been obtained in our fits over the entire range of Q$^2$ and $x$ in which covered by the data. Within this range, it is observed that the Jacobi polynomials approach are more consistent with other methods in the literature. In this paper, the small-$x$ behavior of polarized gluon distribution $x \Delta g(x,Q^2)$ is examined by using the recent DIS data which cover a very low values of $x$ especially the very recent high-precision measurements from {\tt COMPASS16}. The striking feature of obtained polarized gluon distribution is its positivity throughout and clearly away from zero in the regime $x \gtrsim 0.0001$. Overall, we see that the improvement in the determination of the polarized gluon distribution at NNLO approximation is minor due to the lack of polarized DIS data which cover a wide range of $x$, especially at small $x$. However, there is some indication that the biggest change in going from NLO to NNLO is in the polarized gluon distribution. For total quark and gluon polarized distributions the obtained uncertainties are slightly smaller, and it must be remembered that we use the higher order corrections and more constraining data for $10^{-3} < x < 10^{-2}$. A complete understanding of the origin of the proton spin is still lacking and the uncertainties of the polarized gluon PDF at low value of $x$ remain large compared to the currently probed region.
In the future, the current analysis will be extended to include the target mass corrections and higher twist effects. The semi-inclusive DIS asymmetries also can be included which can place constraints on the sea quark polarization.

%
%
\section*{Acknowledgments}
We acknowledge helpful conversations and useful discussions with A. Vogt. The authors also would like to thank E. Leader for carefully reading the manuscript and fruitful discussion and critical remarks.  We are also grateful to S. Taheri Monfared and F. Arbabifar for useful comments and for providing us with the best fit of {\tt THK14} and {\tt AKS14} polarized PDFs.
F. Taghavi-Shahri acknowledges Ferdowsi University of Mashhad for provided facilities to do this research. This work is supported by Ferdowsi University of Mashhad under Grant No 2/39016, 1394/09/25.
H. Khanpour is indebted the University of Science and Technology of Mazandaran and the School of Particles and Accelerators, Institute for Research in Fundamental Sciences (IPM), to support financially this project.

%
%
\section*{Appendix A: FORTRAN package of our NLO and NNLO polarized PDFs}\label{AppendixA}
A \texttt{FORTRAN} package containing our polarized PDFs and their uncertainties at NLO and NNLO approximation as well as the polarized structure functions $x g_1(x,Q^2)$ for the proton, neutron and deuteron can be obtained via Email from the authors upon request. This package includes an example program to illustrate the use of the routines.


%

\begin{thebibliography}{}





\bibitem{Nocera:2014gqa}
E.~R.~Nocera {\it et al.} [NNPDF Collaboration],
Nucl.\ Phys.\ B {\bf 887}, 276 (2014).


\bibitem{Nocera:2014uea}
E.~R.~Nocera,
Phys.\ Lett.\ B {\bf 742}, 117 (2015).


\bibitem{Jimenez-Delgado:2014xza}
P.~Jimenez-Delgado {\it et al.} [Jefferson Lab Angular Momentum (JAM) Collaboration],
Phys.\ Lett.\ B {\bf 738}, 263 (2014).


\bibitem{Jimenez-Delgado:2013boa}
P.~Jimenez-Delgado, A.~Accardi and W.~Melnitchouk,
Phys.\ Rev.\ D {\bf 89}, no. 3, 034025 (2014).


\bibitem{Arbabifar:2013tma}
F.~Arbabifar, A.~N.~Khorramian and M.~Soleymaninia,
Phys.\ Rev.\ D {\bf 89}, no. 3, 034006 (2014).



\bibitem{Monfared:2014nta}
S.~Taheri Monfared, Z.~Haddadi and A.~N.~Khorramian,
Phys.\ Rev.\ D {\bf 89}, no. 7, 074052 (2014)
[Phys.\ Rev.\ D {\bf 89}, no. 11, 119901 (2014)].



\bibitem{Borah:2014esa}
N.~N.~K.~Borah and D.~K.~Choudhury,
Adv.\ High Energy Phys.\  {\bf 2014}, 1 (2014).



\bibitem{Khorramian:2010qa}
A.~N.~Khorramian, S.~Atashbar Tehrani, S.~Taheri Monfared, F.~Arbabifar and F.~I.~Olness,
Phys.\ Rev.\ D {\bf 83}, 054017 (2011).



\bibitem{Blumlein:2010rn}
J.~Blumlein and H.~Bottcher,
Nucl.\ Phys.\ B {\bf 841}, 205 (2010).

\bibitem{Leader:2014uua}
  E.~Leader, A.~V.~Sidorov and D.~B.~Stamenov,
  Phys.\ Rev.\ D {\bf 91}, no. 5, 054017 (2015).


\bibitem{Aschenauer:2015ata} 
E.~C.~Aschenauer, R.~Sassot and M.~Stratmann,
Phys.\ Rev.\ D {\bf 92}, no. 9, 094030 (2015).



\bibitem{Ashman:1987hv}
J.~Ashman {\it et al.} [European Muon Collaboration],
Phys.\ Lett.\ B {\bf 206}, 364 (1988).


\bibitem{Ashman:1989ig}
J.~Ashman {\it et al.} [European Muon Collaboration],
Nucl.\ Phys.\ B {\bf 328}, 1 (1989).




\bibitem{Adolph:2015saz}
C.~Adolph {\it et al.} [COMPASS Collaboration],
Phys.\ Lett.\ B {\bf 753}, 18 (2016).


\bibitem{Alekseev:2007vi}
COMPASS Collaboration, M. Alekseev et~al.,
\newblock Phys.Lett. B660, 458 (2008) .

\bibitem{Alekseev:2009ab}
COMPASS Collaboration, M. Alekseev et~al.,
\newblock Eur.Phys.J. C64 , 171 (2009).


\bibitem{Adolph:2012vj}
COMPASS Collaboration, C. Adolph et~al.,
\newblock Phys.Lett. B718, 922 (2013).

\bibitem{Adolph:2012ca}
COMPASS Collaboration, C. Adolph et~al.,
\newblock Phys.Rev. D87, 052018 (2013).



\bibitem{Airapetian:2004zf}
HERMES Collaboration, A. Airapetian et~al.,
\newblock Phys.Rev. D71, 012003 (2005)


\bibitem{Airapetian:2006vy}
A.~Airapetian {\it et al.} [HERMES Collaboration],
Phys.\ Rev.\ D {\bf 75}, 012007 (2007).

\bibitem{Airapetian:2010ac}
HERMES Collaboration, A. Airapetian et~al.,
\newblock JHEP 1008,  130 (2010).



\bibitem{Adare:2008qb}
PHENIX Collaboration, A. Adare et~al.,
\newblock Phys.Rev. D79, 012003 (2009).

\bibitem{Adare:2008aa}
PHENIX Collaboration, A. Adare et~al.,
\newblock Phys.Rev.Lett. 103, 012003 (2009).


\bibitem{Adare:2014hsq}
A.~Adare {\it et al.} [PHENIX Collaboration],
Phys.\ Rev.\ D {\bf 90}, no. 1, 012007 (2014).

\bibitem{Adare:2010cc}
PHENIX Collaboration, A. Adare et~al.,
\newblock Phys.Rev. D84, 012006 (2011).


\bibitem{Adare:2010xa}
PHENIX Collaboration, A. Adare et~al.,
\newblock Phys.Rev.Lett. 106, 062001 (2011).


\bibitem{Adamczyk:2013yvv}
STAR Collaboration, L. Adamczyk et~al.,
\newblock Phys.Rev. D89, 012001 (2014).

\bibitem{Adamczyk:2012qj}
STAR Collaboration, L. Adamczyk et~al.,
\newblock Phys.Rev. D86,  032006 (2012).

\bibitem{Adamczyk:2014ozi}
L.~Adamczyk {\it et al.} [STAR Collaboration],
Phys.\ Rev.\ Lett.\  {\bf 115}, no. 9, 092002 (2015).

\bibitem{Aggarwal:2010vc}
STAR Collaboration, M. Aggarwal et~al.,
\newblock Phys.Rev.Lett. 106,  062002 (2011).


\bibitem{Adamczyk:2014xyw}
L.~Adamczyk {\it et al.} [STAR Collaboration],
Phys.\ Rev.\ Lett.\  {\bf 113}, 072301 (2014).


\bibitem{Bluemlein:2002be}
J.~Blumlein and H.~Bottcher,
Nucl.\ Phys.\ B \textbf{636} , 225 (2002).


\bibitem{Gluck:2000dy}
M.~Gluck, E.~Reya, M.~Stratmann and W.~Vogelsang,
Phys.\ Rev.\ D \textbf{63}, 094005 (2001).



\bibitem{Leader:2005ci}
E.~Leader, A.~V.~Sidorov and D.~B.~Stamenov,
Phys.\ Rev.\ D \textbf{73} , 034023 (2006).


\bibitem{Leader:2006xc} 
E.~Leader, A.~V.~Sidorov and D.~B.~Stamenov,
Phys.\ Rev.\ D {\bf 75}, 074027 (2007)



\bibitem{deFlorian:2005mw}
D.~de Florian, G.~A.~Navarro and R.~Sassot,
Phys.\ Rev.\ D \textbf{71}, 094018  (2005) .


\bibitem{Goto:1999by}
Y.~Goto \textit{et al.} {[}Asymmetry Analysis
collaboration{]}, 
Phys.\ Rev.\ D \textbf{62} , 034017 (2000),
M.~Hirai, S.~Kumano and N.~Saito {[}Asymmetry Analysis Collaboration{]},
Phys.\ Rev.\ D \textbf{69} , 054021 (2004).




\bibitem{Hirai:2008aj}
M.~Hirai and S.~Kumano {[}Asymmetry Analysis
Collaboration{]},
Nucl.\ Phys.\ B \textbf{813}, 106 (2009).



\bibitem{deFlorian:2008mr}
D.~de Florian, R.~Sassot, M.~Stratmann
and W.~Vogelsang,
Phys.\ Rev.\ Lett.\ \textbf{101}, 072001 (2008).


\bibitem{deFlorian:2009vb}
D.~de Florian, R.~Sassot, M.~Stratmann and W.~Vogelsang,
Phys.\ Rev.\ D {\bf 80}, 034030 (2009).



\bibitem{Moch:2014sna}
S.~Moch, J.~A.~M.~Vermaseren and A.~Vogt,
Nucl.\ Phys.\ B {\bf 889}, 351 (2014)



\bibitem{Cafarella:2005zj}
A.~Cafarella, C.~Coriano and M.~Guzzi,
Nucl.\ Phys.\ B {\bf 748}, 253 (2006).



\bibitem{Lampe:1998eu}
  B.~Lampe and E.~Reya,
  Phys.\ Rept.\  {\bf 332}, 1 (2000).

\bibitem{Zijlstra:1993sh}
  E.~B.~Zijlstra and W.~L.~van Neerven,
  Nucl.\ Phys.\ B {\bf 417}, 61 (1994)
  [Nucl.\ Phys.\ B {\bf 426}, 245 (1994)]
  [Nucl.\ Phys.\ B {\bf 773}, 105 (2007)].


\bibitem{Khorramian:2009xz}
A.~N.~Khorramian, H.~Khanpour and S.~A.~Tehrani,
Phys.\ Rev.\ D {\bf 81}, 014013 (2010).



\bibitem{Khorramian:2008yh}
A.~N.~Khorramian and S.~A.~Tehrani,
Phys.\ Rev.\ D {\bf 78}, 074019 (2008).


\bibitem{AtashbarTehrani:2007odq}
S.~Atashbar Tehrani and A.~N.~Khorramian,
JHEP {\bf 0707}, 048 (2007).


\bibitem{Barker:1982rv}
I.~S.~Barker, B.~R.~Martin and G.~Shaw,
Z.\ Phys.\ C {\bf 19}, 147 (1983).

\bibitem{Barker:1983iy}
I.~S.~Barker and B.~R.~Martin,
Z.\ Phys.\ C {\bf 24}, 255 (1984).

\bibitem{Krivokhizhin:1987rz}
V.~G.~Krivokhizhin, S.~P.~Kurlovich, V.~V.~Sanadze, I.~A.~Savin, A.~V.~Sidorov and N.~B.~Skachkov,
Z.\ Phys.\ C {\bf 36}, 51 (1987).


\bibitem{Krivokhizhin:1990ct}
V.~G.~Krivokhizhin, S.~P.~Kurlovich, R.~Lednicky, S.~Nemecek, V.~V.~Sanadze, I.~A.~Savin, A.~V.~Sidorov and N.~B.~Skachkov,
Z.\ Phys.\ C {\bf 48}, 347 (1990).

\bibitem{Chyla:1986eb}
J.~Chyla and J.~Rames,
Z.\ Phys.\ C {\bf 31}, 151 (1986).

\bibitem{Barker:1980wu}
I.~S.~Barker, C.~S.~Langensiepen and G.~Shaw,
Nucl.\ Phys.\ B {\bf 186}, 61 (1981).



\bibitem{Kataev:1997nc}
A.~L.~Kataev, A.~V.~Kotikov, G.~Parente and A.~V.~Sidorov,
Phys.\ Lett.\ B {\bf 417}, 374 (1998).



\bibitem{Alekhin:1998df}
S.~I.~Alekhin and A.~L.~Kataev,
Phys.\ Lett.\ B {\bf 452}, 402 (1999).




\bibitem{Kataev:1999bp}
A.~L.~Kataev, G.~Parente and A.~V.~Sidorov,
Nucl.\ Phys.\ B {\bf 573}, 405 (2000).



\bibitem{Kataev:2001kk}
A.~L.~Kataev, G.~Parente and A.~V.~Sidorov,
Phys.\ Part.\ Nucl.\  {\bf 34}, 20 (2003)
[Fiz.\ Elem.\ Chast.\ Atom.\ Yadra {\bf 34}, 43 (2003)]
[Phys.\ Part.\ Nucl.\  {\bf 38}, no. 6, 827 (2007)].



\bibitem{Kataev:2005ci}
A.~L.~Kataev,
JETP Lett.\  {\bf 81}, 608 (2005)
[Pisma Zh.\ Eksp.\ Teor.\ Fiz.\  {\bf 81}, 744 (2005)].

\bibitem{Leader:1997kw} 
E.~Leader, A.~V.~Sidorov and D.~B.~Stamenov,
Int.\ J.\ Mod.\ Phys.\ A {\bf 13}, 5573 (1998).

\bibitem{Agashe:2014kda}
K.~A.~Olive {\it et al.} [Particle Data Group Collaboration],
Chin.\ Phys.\ C {\bf 38}, 090001 (2014).




\bibitem{Dokshitzer:1977sg}
Y.~L.~Dokshitzer,
Sov.\ Phys.\ JETP {\bf 46}, 641 (1977)
[Zh.\ Eksp.\ Teor.\ Fiz.\  {\bf 73}, 1216 (1977)].



\bibitem{Gribov:1972ri}
V.~N.~Gribov and L.~N.~Lipatov,
Sov.\ J.\ Nucl.\ Phys.\  {\bf 15}, 438 (1972)
[Yad.\ Fiz.\  {\bf 15}, 781 (1972)].



\bibitem{Lipatov:1974qm}
L.~N.~Lipatov,
Sov.\ J.\ Nucl.\ Phys.\  {\bf 20}, 94 (1975)
[Yad.\ Fiz.\  {\bf 20}, 181 (1974)].



\bibitem{Altarelli:1977zs}
G.~Altarelli and G.~Parisi,
Nucl.\ Phys.\ B {\bf 126}, 298 (1977).




\bibitem{Abe:1998wq}
K.~Abe {\it et al.} [E143 Collaboration],
Phys.\ Rev.\ D {\bf 58}, 112003 (1998).



\bibitem{HERM98}
A.~Airapetian \textit{et al.} {[}HERMES Collaboration{]},
Phys.\ Lett.\ B \textbf{442} (1998) 484.



\bibitem{Adeva:1998vv}
B.~Adeva {\it et al.} [Spin Muon Collaboration],
Phys.\ Rev.\ D {\bf 58}, 112001 (1998).



\bibitem{EMCp} 
J.~Ashman \textit{et al.} {[}European Muon Collaboration{]},
Phys.\ Lett.\ B \textbf{206} (1988) 364; 
J.~Ashman \textit{et al.} {[}European Muon Collaboration{]},
Nucl.\ Phys.\ B \textbf{328} , 1 (1989). 

\bibitem{E155p} 
P.~L.~Anthony \textit{et al.} {[}E155 Collaboration{]},
Phys.\ Lett.\ B \textbf{493} , 19 (2000).

\bibitem{HERMpd} 
A.~Airapetian \textit{et al.} {[}HERMES Collaboration{]},
Phys.\ Rev.\ D \textbf{75} , 012007 (2007).

\bibitem{COMP1} 
M.~G.~Alekseev \textit{et al.} {[}COMPASS Collaboration{]},
Phys.\ Lett.\ B \textbf{690}, 466 (2010).
V.~Y.~Alexakhin \textit{et al.} {[}COMPASS Collaboration{]},
Phys.\ Lett.\ B \textbf{647} , 8 (2007).

\bibitem{E155d} 
P.~L.~Anthony \textit{et al.} {[}E155 Collaboration{]},
Phys.\ Lett.\ B \textbf{463} , 339 (1999).

\bibitem{COMP2005} 
E.~S.~Ageev {\it et al.}  [COMPASS Collaboration],
Phys.\ Lett.\ B {\bf 612}, 154 (2005).


\bibitem{COMP2006} 
V.~Y.~.Alexakhin {\it et al.}  [COMPASS Collaboration],
Phys.\ Lett.\ B {\bf 647}, 8 (2007).


\bibitem{E142n} 
P.~L.~Anthony \textit{et al.} {[}E142 Collaboration{]},
Phys.\ Rev.\ D \textbf{54} , 6620 (1996).

\bibitem{E154n} 
K.~Abe \textit{et al.} {[}E154 Collaboration{]},
Phys.\ Rev.\ Lett.\ \textbf{79} , 26 (1997).

\bibitem{HERMn} 
K.~Ackerstaff \textit{et al.} {[}HERMES Collaboration{]},
Phys.\ Lett.\ B \textbf{404} , 383 (1997).

\bibitem{JLABn2004} 
X.~Zheng {\it et al.}  [Jefferson Lab Hall A Collaboration],
Phys.\ Rev.\ C {\bf 70}, 065207 (2004).


\bibitem{JLABn2005} 
K.~Kramer, D.~S.~Armstrong, T.~D.~Averett, W.~Bertozzi, S.~Binet, C.~Butuceanu, A.~Camsonne and G.~D.~Cates {\it et al.},
Phys.\ Rev.\ Lett.\  {\bf 95}, 142002 (2005).

\bibitem{Khanpour:2016pph}
H.~Khanpour and S.~A.~Tehrani,
Phys.\ Rev.\ D {\bf 93}, 014026 (2016).


\bibitem{AtashbarTehrani:2012xh}
S.~Atashbar Tehrani,
Phys.\ Rev.\ C {\bf 86} , 064301 (2012) .


\bibitem{Martin:2002aw}
A.~D.~Martin, R.~G.~Roberts, W.~J.~Stirling and R.~S.~Thorne,
Eur.\ Phys.\ J.\ C {\bf 28}, 455 (2003).

\bibitem{Pumplin:2001ct}
J.~Pumplin, D.~Stump et al., Phys.\ Rev.\ D {\bf 65} , 014013 (2001).


\bibitem{Monfared:2011xf}
S.~T.~Monfared, A.~N.~Khorramian and S.~A.~Tehrani,
J.\ Phys.\ G {\bf 39}, 085009 (2012).

\bibitem{James:1994vla}
F.~James,
``MINUIT Function Minimization and Error Analysis:  Reference Manual Version 94.1,''
CERN-D-506.


\bibitem{Ball:2013tyh} 
R.~D.~Ball {\it et al.} [NNPDF Collaboration],
Phys.\ Lett.\ B {\bf 728}, 524 (2014).


\bibitem{Ball:2013lla} 
R.~D.~Ball {\it et al.} [NNPDF Collaboration],
Nucl.\ Phys.\ B {\bf 874}, 36 (2013).


\bibitem{Sato:2016tuz} 
N.~Sato {\it et al.} [Jefferson Lab Angular Momentum Collaboration],
Phys.\ Rev.\ D {\bf 93}, no. 7, 074005 (2016).


\bibitem{Martin:2003sk} 
A.~D.~Martin, R.~G.~Roberts, W.~J.~Stirling and R.~S.~Thorne,
Eur.\ Phys.\ J.\ C {\bf 35}, 325 (2004).



\bibitem{deFlorian:2011ia}
D.~de Florian, R.~Sassot, M.~Stratmann and W.~Vogelsang,
Prog.\ Part.\ Nucl.\ Phys.\  {\bf 67}, 251 (2012).


\bibitem{deFlorian:2014yva}
D.~de Florian, R.~Sassot, M.~Stratmann and W.~Vogelsang,
Phys.\ Rev.\ Lett.\  {\bf 113}, no. 1, 012001 (2014).



\bibitem{Lacombe:1981eg}
M.~Lacombe, B.~Loiseau, R.~Vinh Mau, J.~Cote, P.~Pires and R.~de Tourreil,
Phys.\ Lett.\ B {\bf 101}, 139 (1981).

\bibitem{Machleidt:1987hj} 
R.~Machleidt, K.~Holinde and C.~Elster,
Phys.\ Rept.\  {\bf 149}, 1 (1987).



\bibitem{Owens:2012bv}
J.~F.~Owens, A.~Accardi and W.~Melnitchouk,
Phys.\ Rev.\ D {\bf 87}, no. 9, 094012 (2013).


\bibitem{Khanpour:2012tk}
H.~Khanpour, A.~N.~Khorramian and S.~A.~Tehrani,
J.\ Phys.\ G {\bf 40}, 045002 (2013).



\bibitem{Blumlein:2006be}
J.~Blümlein, H.~Böttcher and A.~Guffanti,
 Nucl.\ Phys.\ B \textbf{774}, 182 (2007).
 Nucl.\ Phys.\ Proc.\ Suppl.\ \textbf{135}, 152 (2004).



\bibitem{ABKM} 
 S.~Alekhin, J.~Blümlein, S.~Klein and S.~Moch, 
 Phys.\ Rev.\ D \textbf{81}, 014032 (2010).


\bibitem{JR} 
 M.~Glück, E.~Reya and C.~Schuck, 
 Nucl.\ Phys.\ B \textbf{754}, 178 (2006).
 P.~Jimenez-Delgado and E.~Reya, 
 Phys.\ Rev.\ D \textbf{79}, 074023 (2009).


\bibitem{MSTW} 
 A.~D.~Martin, W.~J.~Stirling, R.~S.~Thorne and G.~Watt, 
 Eur.\ Phys.\ J.\ C \textbf{64}, 653 (2009).


\bibitem{H1ZEUS} H1 and ZEUS collab.
 F.~D.~Aaron \textit{et al.} 
 JHEP \textbf{1001}, 109 (2010).


\bibitem{d'Enterria:2015toz}
D.~d'Enterria and P.~Z.~Skands,
arXiv:1512.05194 [hep-ph].


\bibitem{Adare:2015ozj}
A.~Adare {\it et al.} [PHENIX Collaboration],
Phys.\ Rev.\ D {\bf 93}, no. 1, 011501 (2016).



\bibitem{Adare:2015gsd}
A.~Adare {\it et al.} [PHENIX Collaboration],
arXiv:1504.07451 [hep-ex].




\bibitem{Aschenauer:2014cki}
E.~C.~Aschenauer {\it et al.},
arXiv:1409.1633 [physics.acc-ph].


\bibitem{Miller:2015tjf}
G.~A.~Miller, M.~D.~Sievert and R.~Venugopalan,
arXiv:1512.03111 [nucl-th].


\bibitem{Yang:2013yeb}
J.~C.~Yang {\it et al.},
Nucl.\ Instrum.\ Meth.\ B {\bf 317}, 263 (2013).


\bibitem{Accardi:2012qut}
A.~Accardi {\it et al.},
arXiv:1212.1701 [nucl-ex].


\end{thebibliography}
%

\end{document}